%% file: Loss_GCWS_2025.tex
\documentclass[10pt,conference]{IEEEtran}
\IEEEoverridecommandlockouts

\input{macros}

\IEEEaftertitletext{\vspace{-1.5em}}
\begin{document}
\bstctlcite{IEEEexample:BSTcontrol}

\title{Combining Physical Layer Security and Symmetric Encryption to Secure Rate-Splitting Multiple Access}

\author{\IEEEauthorblockN{Bastian Loss, Mohammad Parvini, Ahmad Nimr, and Gerhard Fettweis}
\IEEEauthorblockA{Vodafone Chair Mobile Communications Systems, Technische Universit{\"a}t Dresden, Germany\\
Email: \{bastian.loss, mohammad.parvini, ahmad.nimr, gerhard.fettweis\}@tu-dresden.de}
}
\vspace{-1em}

\maketitle
\input{acronyms}
\input{chapters/00_Abstract}

\input{chapters/01_Introduction}
\input{chapters/02_System_Model}
\input{chapters/03_Problem_Formulation}
\input{chapters/04_Iterative_Algorithm}
\input{chapters/05_Results}
\input{chapters/06_Conclusion}
\input{chapters/08_Acknowledgement}
\vspace{-1pt}

\bibliographystyle{IEEEtran}
\bibliography{IEEEabrv,./Loss_GCWS_2025.bib}
\end{document}

%% file: macros.tex
\usepackage{cite}
\usepackage{eso-pic}
\usepackage{amsmath,amssymb,amsfonts,amsthm}
\usepackage{orcidlink}
\usepackage[linesnumbered,ruled,vlined]{algorithm2e}
\SetNlSty{}{}{}
\SetKwInput{KwInput}{Input}
\SetKwInput{KwOutput}{Output} 
\SetKwComment{Comment}{$\triangleright$\ }{}

\SetCommentSty{fontcommand}
\def\plotLinewidth{0.8pt}
\def\plotMarkersize{2pt}

\def\plotdistRedVert{-8pt}

\newcommand{\extraYSpaceBetweenEquations}{2pt}
\newcommand{\negSpaceFormula}{\! \! \! \! \! \! \! \! }
\newcommand{\negVeryTiny}{\kern -0.5pt}
\newcommand{\extraDistSum}{-2.5pt}

\newlength{\PZEROsubjectToSpaceOptprob}
\newlength{\PZEROconstraintsSpaceOptprob}
\newlength{\subjectToSpaceOptprob}
\newlength{\addLengthSpaceOptProb}
\newlength{\addLengthSpaceOptProbNeg}
\newlength{\hphantomSt}
\newlength{\constraintsSpaceOptprob}
\newlength{\constraintsSpaceSmallerBracketsOptprob}
\usepackage{cancel}
\usepackage{textgreek}
\definecolor{fourClassColour1}{HTML}{8dd3c7}
\definecolor{fourClassColour2}{HTML}{ffffb3}
\definecolor{fourClassColour3}{HTML}{bebada}
\definecolor{fourClassColour4}{HTML}{fb8072}

\definecolor{fourClassColourArrow3}{HTML}{928bc1}
\definecolor{fourClassColourArrow2}{HTML}{cccc00}
\definecolor{fourClassColourArrow4}{HTML}{f94c39}

\definecolor{RMSEColour1}{HTML}{ebb11f}
\definecolor{RMSEColour2}{HTML}{87347d}
\definecolor{RMSEColour3}{HTML}{c92226}
\definecolor{RMSEColour4}{HTML}{5b5c9b}
\definecolor{RMSEColour5}{HTML}{9ec288}
\definecolor{RMSEColour6}{HTML}{a8ad1e}
\definecolor{RMSEColour7}{HTML}{983a1b}
\definecolor{RMSEColour8}{HTML}{00a6a6}
\definecolor{RMSEColour9}{HTML}{d36b0f}
\definecolor{RMSEColourCRB}{HTML}{000000}

\def\tiksizeVarInnerPlot{6}
\def\fontsizevar{9}

\def\fontsizevarsmaller{7.85}
\def\fontsizevarsYLabel{8.5}
\def\fontsizevarsYLabelTC11{8.5}

\usepackage[normalem]{ulem}
\usepackage{tikz}
\usetikzlibrary{positioning, arrows.meta}
\usepackage{contour}
\usepackage[dvipsnames]{xcolor}
\usepackage{pgfplots}
\usepackage[position=top]{subfig}
\usepackage{graphicx}
\usepackage[printonlyused,nohyperlinks, nolist]{acronym}
\usepackage{textcomp}

\usepackage{subcaption}
\usepackage{siunitx}
\usepackage{booktabs}
\usepackage{multirow}
\usepackage{multicol}
\usepackage{enumitem}
\usepackage{tabularx}
\usepackage{bm}
\usepackage{mathtools}
\newcolumntype{L}[1]{>{\raggedright\arraybackslash}p{#1}}
\newcolumntype{C}[1]{>{\centering\arraybackslash}p{#1}}
\newcolumntype{R}[1]{>{\raggedleft\arraybackslash}p{#1}}
\newlist{mylist}{itemize}{1}
\setlist[mylist]{label=\textbullet,
                 nosep, wide, leftmargin=*,
                 before=\vspace{-0.57\baselineskip},
                 after =\vspace{-0.85\baselineskip}}

\def\BibTeX{{\rm B\kern-.05em{\sc i\kern-.025em b}\kern-.08em
    T\kern-.1667em\lower.7ex\hbox{E}\kern-.125emX}}
\hypersetup{
    colorlinks,
    linkcolor={red!50!black},
    citecolor={blue!50!black},
    urlcolor={blue!80!black}
}
\hypersetup{draft}

\DeclareMathOperator{\tr}{tr}

\newcommand{\ma}  [1]{ \mathbf{#1} }

\newcommand{\set} [1]{{\mathcal {#1}}} 

\makeatletter
\newcommand{\subalign}[1]{%
  \vcenter{%
    \Let@ \restore@math@cr \default@tag
    \baselineskip\fontdimen10 \scriptfont\tw@
    \advance\baselineskip\fontdimen12 \scriptfont\tw@
    \lineskip\thr@@\fontdimen8 \scriptfont\thr@@
    \lineskiplimit\lineskip
    \ialign{\hfil$\m@th\scriptstyle##$&$\m@th\scriptstyle{}##$\hfil\crcr
      #1\crcr
    }%
  }%
}
\makeatother

\AddToShipoutPictureFG{%
  \AtPageUpperLeft{%
    \raisebox{-7mm}{%
      \makebox[\paperwidth][c]{%
        \parbox{\dimexpr\paperwidth-4cm\relax}{%
          \centering
          \color{gray}
          \fontsize{7}{7}\selectfont
          This document is a preprint of:
          B.~Loss, M.~Parvini, A.~Nimr et al.,
          ``Combining Physical Layer Security and
          Symmetric Encryption to Secure
          Rate-Splitting Multiple Access,''
          in Proceedings of 2025 Globecom Workshops
          (GC Wkshps 2025), Taipei, Taiwan,
          (p.~6), Dec.~2025.
          DOI: 10.1109/GCWkshps68340.2025.11590874
        }%
      }%
    }%
  }%
}

\AddToShipoutPictureFG{%
  \AtPageLowerLeft{%
    \raisebox{6mm}{%
      \makebox[\paperwidth][c]{%
        \parbox{\dimexpr\paperwidth-4cm\relax}{%
          \centering
          \color{gray}
          \fontsize{7}{7}\selectfont
          \textcopyright~2025 IEEE.
          Personal use of this material is permitted.
          Permission from IEEE must be obtained for
          all other uses, in any current or future
          media, including reprinting/republishing
          this material for advertising or promotional
          purposes, creating new collective works,
          for resale or redistribution to servers or
          lists, or reuse of any copyrighted component
          of this work in other works.
        }%
      }%
    }%
  }%
}

%% file: acronyms.tex
\begin{acronym}
    \acro{SINR}{signal-to-interference-plus-noise ratio}
    \acro{SNR}{signal-to-noise ratio}
    \acro{CSI}{channel state information}
    \acro{DL}{downlink}
    \acro{BS}{base station}
    \acro{EE}{energy efficiency}
    \acro{SE}{spectral efficiency}
    \acro{XR}{extended reality} 
    \acro{MCS}{modulation and coding schemes}
    
    \acro{MIMO}{multiple-input multiple-output}
    \acro{MU-MIMO}{multi-user multiple-input multiple-output}
    \acro{MISO}{multiple-input single-output}
    \acro{SISO}{single-input single-output}
	\acro{BC}{broadcast channel}
  
    \acro{NOMA}{non-orthogonal multiple access}
    \acro{SDMA}{space division multiple access}
    \acro{MU-LP}{multi-user linear precoding}
    \acro{RS}{rate-splitting}
    \acro{RSMA}{rate-splitting multiple access} 
    \acro{SIC}{successive interference cancellation}

    \acro{QoS}{quality of service}
    \acro{WSR}{weighted sum-rate}
    \acro{UB}{upper bound}
    \acro{SCA}{successive convex approximation}
    
    \acro{PLS}{physical layer security}
    \acro{ED}{eavesdropper}
    \acroplural{ED}[EDs]{eavesdroppers}  
    \acro{FSDT}{fully secure data transmission}    
\end{acronym}

%% file: chapters/00_Abstract.tex
\begin{abstract}\label{abstract} 
The demand for increased \ac{SE} in downlink \ac{MU-MIMO} systems has led to the development of various multiple access techniques. Among these, \ac{RSMA} can improve \ac{SE} over \ac{SDMA} and \ac{NOMA} by splitting each user's data into common and private parts. The common parts from all users are then encoded into a single common message, which is transmitted alongside the private messages using an \ac{RSMA}-specific precoder. Recent results showed that the precoder can be modified to enable \ac{PLS} by integrating secrecy rate constraints into the design. However, this approach only secures the private messages, while the common part of each user remains unprotected. To address this limitation, we propose a hybrid-\ac{PLS} \ac{RSMA} scheme that symmetrically encrypts all user data and embeds key information within the \ac{PLS}-secured private messages. Additionally, the \ac{RSMA} precoder is adapted to ensure that the private messages are sufficiently sized to carry the key information. The precoder design is formulated as a non-convex \ac{SE} maximization problem and solved using \ac{SCA}. Analytical and numerical evaluations demonstrate that the proposed hybrid-\ac{PLS} \ac{RSMA} can significantly improve \ac{SE} compared to partially secure \ac{RSMA} and secure \ac{SDMA}, when complete user data security is required. \looseness=-1 
\end{abstract}
\acresetall

%% file: chapters/01_Introduction.tex
\vspace{-1em}
\section{Introduction}
\IEEEPARstart{F}{\lowercase{uture}} mobile communication systems require high \ac{SE} to accommodate the growing number of devices and the rising demand from data-intensive applications such as \ac{XR} or edge computing. \Ac{MU-MIMO} techniques address this by serving multiple users on the same time-frequency resource by leveraging the spatial dimension through multiple antennas. 
Widely studied \ac{MU-MIMO} methods include \ac{SDMA} and \ac{NOMA}, which adopt different strategies for handling interference. In \ac{SDMA}, the receiver treats any interference as noise, whereas \ac{NOMA} uses superposition coding at the transmitter and \ac{SIC} at the receiver to sequentially decode and subtract interference. In terms of performance, either \ac{SDMA} or \ac{NOMA} may provide higher \ac{SE} depending on factors such as user count, antenna configuration or channel conditions~\cite{Mao2022}.\looseness=-1

To further enhance \ac{SE}, the recently proposed \ac{RSMA} framework~\cite{Mao2022} integrates both strategies within several unified architectures, such as downlink linearly precoded 1-layer \ac{RSMA} (referred to simply as \ac{RSMA} hereafter). At the transmitter, \ac{RSMA} splits the message of each user into a common part and a private part. The common parts from all users are then combined into a single common message, which is jointly precoded and transmitted alongside the private messages.
At the receiver, each user first decodes and subtracts the common message via \ac{SIC} to reduce interference. The private message is then decoded while treating any remaining interference as noise. This way, \ac{RSMA} unifies \ac{SDMA} and \ac{NOMA} to achieve equal or higher \ac{SE} in all scenarios~\cite{Mao2022}. \looseness=-1

Another key aspect of future communication systems is \ac{PLS}, which prevents unauthorized decoding by degrading an attacker’s physical channel \cite{Wyner1975}. This idea can also be applied to \ac{RSMA}, as demonstrated in~\cite{Lei2024}, where a secure \ac{RSMA} mmWave system was proposed to enhance confidentiality against an external \ac{ED} with unknown \ac{CSI}. This system transmits the common message on spatial paths overlapping between users, while private messages use non-overlapping paths. Consequently, security degrades if an attacker also overlaps with one or more of these paths, making it strongly dependent on the positions of the users and the \ac{ED}.
An alternative strategy is to introduce additional noise to degrade the channel of an external attacker. To achieve this, \cite{Fu2020} utilizes imperfect \ac{CSI} to construct a precoder that prevents the \ac{ED} from decoding the common message. As a result, the common message acts as additional noise to the \ac{ED} when decoding private messages. Alternatively, the authors of~\cite{Cai2021,Xia2023ANApproach} introduce artificial noise into their transmitter design, considering scenarios without \ac{CSI}~\cite{Cai2021} and with perfect \ac{CSI}~\cite{Xia2023ANApproach}. However, these methods cannot prevent internal eavesdropping by legitimate users, as introducing additional noise may interfere with them decoding their intended messages. To address this, \cite{Xia2022} proposes an \ac{RSMA} precoder utilizing perfect \ac{CSI} to prevent users from decoding other users’ private messages. An important limitation of this approach is that the common message cannot be secured, as it must remain decodable by all users. Thus, only the private messages can be utilized for confidential data transfer, effectively reducing \ac{RSMA} to \ac{SDMA}. As a result, the achievable \ac{SE} decreases in all scenarios where \ac{RSMA} outperforms \ac{SDMA}, particularly in the region of operation of \ac{NOMA} ~\cite{Mao2022}.\looseness=-1

\begin{figure*}[!t]
    \centering
    \captionsetup{justification=centering,labelsep=space}
    \includegraphics[
        width=\textwidth
    ]{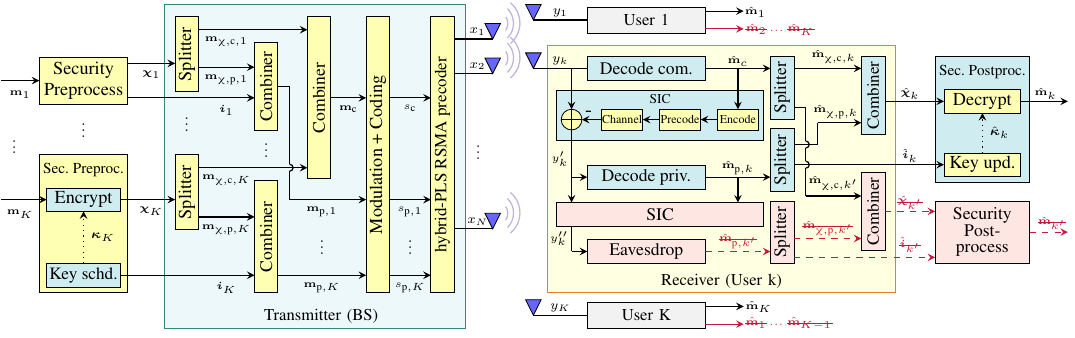}
    \caption{System model for downlink linearly precoded 1-layer hybrid-\ac{PLS} \ac{RSMA}}
    \label{fig:RSMALayerEncFigure}
    \vspace{-1.7em}
\end{figure*}

To address this limitation, we propose a hybrid-\ac{PLS} \ac{RSMA} scheme that combines symmetric encryption with secure \ac{RSMA} precoding. In our approach, all user data is symmetrically encrypted, while the key information is transmitted exclusively within the \ac{PLS}-secured private messages, ensuring only the intended user can decode it. This enables the inclusion of encrypted data in both common and private messages without compromising security. For this purpose, we propose a modified \ac{RSMA} precoder that ensures the required size as well as the security of the private messages. \looseness=-1

The remainder of this paper is organized as follows. Section~\ref{sec:sysmodel} introduces the system model, while Section~\ref{sec:probform} casts the \ac{RSMA} precoder design as a non-convex \ac{SE} maximization problem and reformulates it using \ac{SCA}. Section~\ref{sec:iteralgo} then presents the \ac{SCA}-based algorithm for obtaining a locally optimal solution. Thereafter, Section~\ref{sec:perfRes} reports analytical and numerical results comparing the proposed hybrid-\ac{PLS} \ac{RSMA} with secure \ac{RSMA}\cite{Xia2022} and secure \ac{SDMA}. Finally, Section~\ref{sec:concl} concludes this work. 

%% file: chapters/02_System_Model.tex
\section{System Model}
\label{sec:sysmodel}
For this work we consider a downlink \ac{MU-MIMO} system employing a \ac{BS} with $N$ antennas that provides confidential multiple access to $k \in  \mathcal{K} = \{1,\ldots, K\}$ single-antenna users. It is assumed that the \ac{BS} and all users have access to perfect \ac{CSI} and use the same family of \ac{MCS}. The users are served using hybrid-\ac{PLS} \ac{RSMA}, which consists of security-related pre- and postprocessing steps in combination with an adjusted \ac{RSMA} transmitter and receiver. The system is illustrated in Fig. \ref{fig:RSMALayerEncFigure}.\looseness=-1

\subsection{Security Preprocessing and Postprocessing}
\label{sec:secprepostproc}
The security preprocessing stage employs symmetric encryption to transform the message $\mathbf{m}_k$ of each user $k$ into a ciphertext $\boldsymbol{\chi}_k$, utilizing a symmetric encryption key $\boldsymbol{\kappa}_k$. 
Instead of transmitting a unique full-length key ${\bar{\boldsymbol{\kappa}}}_k$ for each $\mathbf{m}_k$, an abstract key schedule is introduced to include compact key update information $\boldsymbol{i}_k$ alongside each ciphertext. The \ac{RSMA} system then securely transmits $\boldsymbol{i}_k$ by utilizing \ac{PLS}, while $\boldsymbol{\chi}_k$ is transmitted without security considerations. 

At the receiver, security postprocessing begins with updating the symmetric decryption key based on the decoded key information $\hat{\boldsymbol{i}}_k$. Subsequently, the updated key ${\hat{\boldsymbol{\kappa}}}_k$ is used to decrypt the received ciphertext $\hat{\boldsymbol{\chi}}_k$ into the plaintext message $\hat{\mathbf{m}}_k$, while other ciphertexts $\hat{\boldsymbol{\chi}}_{k'}$ intended for users $k' \in \mathcal{K} \setminus \{k\}$ remain indecipherable to user $k$.
This follows because the required $\hat{\boldsymbol{\kappa}}_{k'}$ cannot be reconstructed without $\hat{\boldsymbol{i}}_{k'}$, which is assumed to be protected by \ac{PLS} and hence inaccessible to user $k$. \looseness=-1

A simple example of a key update mechanism involves dividing a new key $\overline{\boldsymbol{\kappa}}_k$ into $T$ smaller key update fragments $[\boldsymbol{i}^{1}_k, \boldsymbol{i}^{2}_k, \ldots, \boldsymbol{i}^{T}_k]$. After receiving all $T$ fragments, the receiver reconstructs the new key $\overline{\boldsymbol{\kappa}}_k$ and updates its decryption key as ${\hat{\boldsymbol{\kappa}}}_k \leftarrow \overline{\boldsymbol{\kappa}}_k$. Note that it is sufficient to update symmetric encryption keys at intervals of multiple hours \cite{nistsp80057Part1Rev5} without compromising security, thus each key fragment $\boldsymbol{i}_k^t$ can be limited to a few bits.

\subsection{Transmit and Receive Procedure}
As shown in Fig. \ref{fig:RSMALayerEncFigure}, the proposed \ac{RSMA} transmitter accepts the ciphertext $\boldsymbol{\chi}_{k}$ and the key information $\boldsymbol{i}_k$ of each user $k$. The rate-splitting technique is then applied to all ciphertexts $\boldsymbol{\chi}_{1} \ldots \boldsymbol{\chi}_{K}$, to produce the respective common and private ciphertext parts $\mathbf{m}_{\text{\textchi},\text{c},k}$ and $\mathbf{m}_{\text{\textchi},\text{p},k}$ of each user $k$. The private message $\mathbf{m}_{\text{p},k}$ of user $k$ is then constructed by combining $\mathbf{m}_{\text{\textchi},\text{p},k}$ with the key information $\boldsymbol{i}_k$. Similarly, the overall common message $\mathbf{m}_{\text{c}}$ is constructed by concatenating the common message parts of all users $\mathbf{m}_{\text{c}} = [\mathbf{m}_{\text{\textchi},\text{c},1},\ldots,\mathbf{m}_{\text{\textchi},\text{c},K}]$. In the next step, channel coding and modulation are applied to the common and private messages $\mathbf{m}_{\text{c}}, \mathbf{m}_{\text{p}, 1}, \ldots, \mathbf{m}_{\text{p}, K}$, to produce the respective common $s_{\text{c}}$ and private streams $s_{\text{p},k}$. Note that $\mathbf{s} = [s_{\text{c}}, \, s_{\text{p},1}, \, \ldots, \, s_{\text{p},K}]^T$ denotes the symbol vector for a single channel use, with source signals that are assumed to be uncorrelated and of unit power such that $\mathbb{E}[\mathbf{s}\mathbf{s}^H]= \mathbf{I}$. Finally, the security-aware \ac{RSMA} precoders $\mathbf{w}_{\text{c}}, \mathbf{w}_{\text{p},k} \in \mathbb{C}^{N \times 1}$ are applied to the common and private streams in order to generate the transmit signal $\mathbf{x} \in \mathbb{C}^{N \times 1}$: \looseness=-1
\begin{equation} \small
    \ma{x} = \ma{W} \ma{s} = \mathbf{w}_{\text{c}} \, s_{\text{c}} +  \sum_{k\in \mathcal{K}} \mathbf{w}_{\text{p},k} \, s_{\text{p},k},
\end{equation}	
where $\mathbf{W} = [\mathbf{w}_{\text{c}}, \, \mathbf{w}_{\text{p},1},  \, \ldots, \, \mathbf{w}_{\text{p},K}]$ describes the precoder matrix. Each user $k$ then receives the signal $y_{k} \in \mathbb{C}$:\looseness = -1
\begin{equation}\small
    y_k = \mathbf{h}^H_k \mathbf{x} + n_k,
\end{equation}
where $\mathbf{h}_k \in \mathbb{C}^{N \times 1}$ denotes the flat-fading channel realization between the \ac{BS} and user $k$, while $n_k \sim \mathcal{CN}(0, \sigma_{n,k}^2) $ refers to the receiver noise of user $k$, with $\sigma_{\text{n},k}^2 = \sigma_{\text{n}}^2, \,\forall k \, \in \, \mathcal{K}$. 

Upon reception, each user $k$ decodes the common message $\hat{\mathbf{m}}_{\text{c}}$ from $y_k$ and extracts its respective partial ciphertext $\hat{\mathbf{m}}_{\text{\textchi},\text{c},k}$. Thereafter, user $k$ re-encodes, precodes, and re-applies the channel $\mathbf{h}_k^H$ to $\hat{\mathbf{m}}_{\text{c}}$ in order to perform perfect \ac{SIC} on the received signal $y_k$. This yields the intermediate signal $y'_{k} = y_k - \mathbf{h}^H_k \mathbf{w}_{\text{c}} \hat{\mathbf{m}}_{\text{c}}$ from which user $k$ decodes the private message $\hat{\mathbf{m}}_{\text{p},k}$. The private message $\hat{\mathbf{m}}_{\text{p},k}$ is then split into the partial ciphertext $\hat{\mathbf{m}}_{\text{\textchi},\text{p},k}$ and the key information $\hat{\boldsymbol{i}}_{k}$. The full ciphertext $\hat{\boldsymbol{\chi}}_{k}$ is then reconstructed by combining both partial ciphertexts $\hat{\mathbf{m}}_{\text{\textchi},\text{c},k}$ and $\hat{\mathbf{m}}_{\text{\textchi},\text{p},k}$. Finally, user $k$ can pass $\hat{\boldsymbol{\chi}}_k$ and $\hat{\boldsymbol{i}}_{k}$ to the security postprocessing, to obtain the message $\ma{\hat{m}}_k$. \looseness=-1

After decoding its own message $\hat{\mathbf{m}}_k$, user $k$ may attempt to eavesdrop on any other user $k' \in \mathcal{K} \setminus \{k\}$, by decoding their message $\hat{\mathbf{m}}_{k'}$. In order to accomplish this, user $k$ must decode both the ciphertext $\hat{\boldsymbol{\chi}}_{k'}$ intended for user $k'$ and all associated key information $\hat{\boldsymbol{i}}_{k'}$. 
The \ac{RSMA} transmission system inherently reveals the common part of the ciphertext $\hat{\mathbf{m}}_{\text{\textchi},\text{c},k'}$ as it is part of the shared common message $\hat{\mathbf{m}}_c$.
However, the remaining ciphertext $\hat{\mathbf{m}}_{\text{\textchi},\text{p},k'}$ and the key information $\hat{\boldsymbol{i}}_{k'}$ are transmitted in the \ac{PLS}-secured private message $\hat{\mathbf{m}}_{\text{p},k'}$, which user $k$ can not decode. Hence, user $k$ is unable to reconstruct $\hat{\mathbf{m}}_{k'}$. \looseness=-1

\subsection{Achievable Rates}
Based on these assumptions, we denote $R_{\text{c},k}$ and $R_{\text{p},k}$ as the maximum achievable rates for user $k$ when decoding the common and private streams, respectively:

\vspace{-1em}
\begin{subequations}
    \begin{align}
       		R_{\text{c},k} & = \log_2\left(1 + \frac{|\mathbf{h}_{k}^H \mathbf{w}_{\text{c}}|^2 }{\sum_{j\in \mathcal{K}}  |\mathbf{h}_{k}^H \mathbf{w}_{\text{p},j}|^2 + \sigma_{\text{n}}^2} \right) \label{eq:R_k_c_term}, \\
            R_{\text{p},k} & = \log_2\left(1 + \frac{|\mathbf{h}_{k}^H \mathbf{w}_{\text{p},k}|^2 }{\sum_{j \in \mathcal{K}\setminus \{k\}} |\mathbf{h}_{k}^H \mathbf{w}_{\text{p},j}|^2  + \sigma_{\text{n}}^2} \right) \label{eq:R_k_p_term}.
    \end{align}
\end{subequations}
We further denote $R_{\text{c}}$ as the maximum transmission rate at which the common stream $s_{\text{c}}$ remains decodable by all users:

\vspace{-0.5em}
\begin{equation} 
    \begin{aligned}
    \label{eq:R_c_term}
       		R_{\text{c}} & = \min_{ k \in \set{K}} \{R_{\text{c},k}\}.
    \end{aligned}
\end{equation}
Based on $R_{\text{c}}$, we also define the common rate allocation vector $\ma{c} = [C_1, \ldots, C_K]$, with $\sum_{j \in K}{C_{j}} \leq R_{\text{c}}$, where $C_k \, \in \, \mathbb{R}^+$ represents the portion of $R_{\text{c}}$ allocated to user $k$. 
Based on these partial rates, we define the total rate $R_{k}$ of each user $k$ as $R_{k} = C_k + R_{\text{p}, k}$. Further, we denote $R_{\text{p}, k',k}$ as the achievable eavesdropping rate at which an unintended user $k' \in \mathcal{K} \setminus \{k\}$ can decode the private stream $s_{\text{p},k}$ of another user $k$. 
\begin{equation}
    \begin{aligned}
        \label{eq:R_kprime_k_p}
            R_{\text{p}, k',k} & = \log_2\left(1 + \frac{|\mathbf{h}_{k'}^H \mathbf{w}_{\text{p}, k}|^2 }{\sum_{j \in \mathcal{K}\setminus \{k', k\}} |\mathbf{h}_{k'}^H \mathbf{w}_{\text{p}, j}|^2  + \sigma_{\text{n}}^2} \right)
    \end{aligned}
\end{equation}
Based on these expressions, we can also denote the secrecy rate $R_{\text{s},k} \, \in \, [0,R_{\text{p},k}]$ of user $k$'s private stream as
\begin{equation} 
    \begin{aligned}
        \label{eq:R_kprime_k_s}
       		R_{\text{s},k} & = \left[ R_{\text{p},k} - \max_{k' \in \set{K} \setminus \{k\}} \{R_{\text{p},k',k}\}\right]^{+},
    \end{aligned}
\end{equation}
where $[x]^{+} = \max\{x,0\}$. In this case, $R_{\text{s}, k}$ can be interpreted as a rate advantage of the legitimate user $k$ over the most capable eavesdropper. The most capable eavesdropper is the user who achieves the maximum eavesdropping rate ($\max_{ k' \in \set{K} \setminus \{k\}} \{R_{\text{p}, k',k}\}$) across all potential eavesdroppers $k'$.

%% file: chapters/03_Problem_Formulation.tex
\section{Problem Formulation and Optimization}
\label{sec:probform}
In this section, we formulate the non-convex optimization problem of maximizing the \ac{SE} of multiple users under the private message size and secrecy rate constraints. To address the non-convexity, we employ the \ac{SCA}-based method from \cite{Xia2022} to transform the problem into a convex form, which is then solved via an iterative algorithm.
\subsection{Problem Formulation}
By utilizing our proposed secure hybrid-\ac{PLS} \ac{RSMA} scheme, the optimization problem can be formulated as
\begin{subequations} 
\label{eq:P0FirstProblem} \small
    \begin{align}
        & \mathrm{P0} \!: \max_{\mathbf{c},\mathbf{W}} \;  \sum_{k \in \set{K}} \left(C_k + R_{\text{p},k}  - R_{\text{i},k}^{\mathrm{req}} \right)
        \label{eq:constC_objFunc} \\
        & \hspace{\PZEROsubjectToSpaceOptprob} \textrm{s.t.} \; R_{\text{s},k} \geq R_{\text{s},k}^{\mathrm{req}} \quad \forall k \in \set{K}, \label{eq:constC_reqSecrate}  \\
        & \hspace{\PZEROconstraintsSpaceOptprob} R_{\text{p},k} \geq R_{\text{i},k}^{\mathrm{req}} \quad \forall k \in \set{K}, \label{eq:constC_reqBaseRate} \\
        & \hspace{\PZEROconstraintsSpaceOptprob} \sum_{j \in \set{K}}{C_{j}} \leq R_{\text{c}}, \label{eq:constC_commonDecodable}\\
        & \hspace{\PZEROconstraintsSpaceOptprob} \mathbf{c} \succeq \mathbf{0},   \label{eq:constC_cNoneg} \\
        & \hspace{\PZEROconstraintsSpaceOptprob} \tr (\mathbf{W} \mathbf{W}^H)  \leq P_{\text{t}},  \label{eq:constC_TXPwr}
    \end{align}
\end{subequations}
where $R_{\text{i},k}^{\mathrm{req}}$ denotes the rate required by user $k$ to transmit the key information $\boldsymbol{i}_k$ and $R_{\text{s},k}^{\mathrm{req}}$ denotes the required secrecy rate. Accordingly, constraint \eqref{eq:constC_reqBaseRate} ensures that the private message of user $k$ can always fit $\boldsymbol{i}_k$. Similarly, constraint \eqref{eq:constC_reqSecrate} ensures that the achieved secrecy rate $R_{\text{s},k}$ for the private message of user $k$ is at least $R_{\text{s},k}^{\mathrm{req}}$. Additionally, constraints \eqref{eq:constC_commonDecodable} and \eqref{eq:constC_cNoneg} ensure that the common stream allocation $\ma{c}$ is feasible by restricting it to be non-negative and not greater than the available common rate $R_{\text{c}}$. Constraint \eqref{eq:constC_TXPwr} limits the maximum transmit power to $P_{\text{t}}$. \looseness=-1
\subsection{Problem Reformulation}
However, the optimization problem $\mathrm{P0}$ is generally non-convex due to the non-convex objective function \eqref{eq:constC_objFunc} and the non-convex constraints \eqref{eq:constC_reqSecrate}-\eqref{eq:constC_commonDecodable}. To address this, we introduce the variables: $\bm{\alpha} = \{\alpha_{\text{c},k}, \; \alpha_{\text{p},k}, \; \alpha_{k',k} \in \mathbb{R}^+ \, | \, \forall k \in \set{K}, k' \in \set{K}\setminus \{k\}\}$, $\bm{\rho} = \{\rho_{\text{c},k}, \; \rho_{\text{p},k}, \; \rho_{k',k} \in \mathbb{R}^+ \,  | \, \forall k \in \set{K}, k' \in \set{K}\setminus \{k\}\}$, and reformulate the problem $\mathrm{P0}$ into $\mathrm{P1}$:
\vspace{-1em}

\allowdisplaybreaks
\begin{subequations} \label{eq:P1firstreform}  \small
    \begin{align}
        & \mathrm{P1} \!: \max_{\mathbf{c},\mathbf{W}, \bm{\alpha}, \bm{\rho}}  \,  \sum_{k \in \set{K}} \left( C_k + \alpha_{\text{p},k}  - R_{\text{i},k}^{\mathrm{req}} \right)\\
        & \hspace{\subjectToSpaceOptprob} \textrm{s.t.} \; \eqref{eq:constC_cNoneg}, \eqref{eq:constC_TXPwr}, \nonumber \\
        & \hspace{\constraintsSpaceSmallerBracketsOptprob} \left[\alpha_{\text{p},k}  - \alpha_{k',k}\right]^{+}  \geq R_{\text{s},k}^{\mathrm{req}}, \forall \; \subalign{&k \in \set{K},\\&k' \in \set{K} \setminus \{k\}}, \label{eq:alphaSecrecyRateDiff}\\
        & \hspace{\constraintsSpaceOptprob} \alpha_{\text{p},k} \geq R_{\text{i},k}^{\mathrm{req}}, \forall k \in \set{K}, \label{eq:alphaBaseRate}  \\
        &  \hspace{\constraintsSpaceOptprob} \sum_{j \in \set{K}}{C_{j}} \leq \alpha_{\text{c},k}, \forall k \in \set{K}, \label{eq:CsumBaseRateConst} \\
        & \hspace{\constraintsSpaceOptprob} 1+\rho_{\text{u},k} \geq 2^{\alpha_{\text{u},k}}, \forall k \in \set{K}, \text{u} \in \{\text{c},\text{p}\},  \label{eq:convex_rhoEavesdrop} \\
        & \hspace{\constraintsSpaceOptprob} 1+\rho_{k',k} \leq 2^{\alpha_{k',k}}, \forall \; \subalign{&k \in \set{K},\\&k' \in \set{K} \setminus \{k\}}, \label{eq:nonconvex_rhoEavesdrop}\\
        & \hspace{\constraintsSpaceOptprob} \frac{|\mathbf{h}_{k}^H \mathbf{w}_{\text{c}}|^2 }{\sum_{j\in \mathcal{K}}  |\mathbf{h}_{k}^H \mathbf{w}_{\text{p},j}|^2 + \sigma_n^2} \geq  \rho_{\text{c},k}, \, \forall k \in \set{K}, \label{eq:nonconvex_commonSINR}\\
        & \hspace{\constraintsSpaceOptprob} \frac{|\mathbf{h}_{k}^H \mathbf{w}_{\text{p},k}|^2 }{\sum_{j \in \mathcal{K}\setminus \{k\}} |\mathbf{h}_{k}^H \mathbf{w}_{\text{p},j}|^2 + \sigma_n^2} \geq \rho_{\text{p},k}, \, \forall k \in \set{K}, \label{eq:nonconvex_privateSINR}\\
        & \hspace{\constraintsSpaceOptprob} \label{eq:nonconvex_eavesdropSINR} \frac{|\mathbf{h}_{k'}^H \mathbf{w}_{\text{p},k}|^2}{\sum_{j \in \mathcal{K}\setminus \{k, k^{'}\}} |\mathbf{h}_{k^{'}}^H \mathbf{w}_{\text{p},j}|^2 + \sigma_n^2} \leq \rho_{k',k}, \forall \, \subalign{&k \in \set{K},\\&k' \in \set{K} \setminus \{k\}}.
    \end{align}
\end{subequations}
In this reformulation, we utilized the fact that the $\max$ operation in constraint \eqref{eq:constC_reqSecrate} can be omitted when the constraint holds for all users $k$ and all eavesdroppers $k'$. A similar simplification was made for constraints \eqref{eq:constC_commonDecodable}. Despite this, problem $\mathrm{P1}$ remains non-convex due to constraints \eqref{eq:nonconvex_rhoEavesdrop}-\eqref{eq:nonconvex_eavesdropSINR}. Hence, we further reformulate constraints \eqref{eq:nonconvex_privateSINR} and \eqref{eq:nonconvex_commonSINR} using a parameter vector $\bm{\beta} = \{\beta_{\text{c},k}, \beta_{\text{p},k} \in \mathbb{R}^+ \, | \, \forall k \in \set{K}\}$: 
\vspace{-1em}

\begin{subequations} \small
    \begin{align}
        &\frac{|\mathbf{h}_{k}^H \mathbf{w}_{\text{u},k}|^2 }{\beta_{\text{u},k}} \geq \rho_{\text{u},k}, \, \forall k \in K, u \in \{\text{c},\text{p}\}, \label{eq:cons_betaRhoCP}\\
        &\sum_{j\in \mathcal{K}}  |\mathbf{h}_{k}^H \mathbf{w}_{\text{p},j}|^2 + \sigma_n^2 \leq \beta_{\text{c},k}, \, \forall k \in K \label{eq:cons_betaCSum},\\
        &\sum_{j \in \mathcal{K}\setminus \{k\}} |\mathbf{h}_{k}^H \mathbf{w}_{\text{p},j}|^2 + \sigma_n^2 \leq \beta_{\text{p},k} , \,  \forall k \in K, \label{eq:cons_betaPSum}
    \end{align}
\end{subequations}
with $\mathbf{w}_{\text{c},k} = \mathbf{w}_{\text{c}}$. This results in the convex constraints \eqref{eq:cons_betaCSum}, \eqref{eq:cons_betaPSum}, and the non-convex constraint \eqref{eq:cons_betaRhoCP}, which cannot be further reformulated and must be linearized.
Therefore, we rewrite constraint \eqref{eq:cons_betaRhoCP} as a function $f(\mathbf{w}_{\text{u},k},\beta_{\text{u},k},\rho_{\text{u},k}) = \frac{|\mathbf{h}_{k}^H \mathbf{w}_{\text{u},k}|^2 }{\beta_{\text{u},k}} - \rho_{\text{u},k}$ and approximate it using the first-order Taylor series for complex vector-valued functions \cite{Hjorungnes2011}. This yields the linearized constraint
\begin{equation} \small
    \frac{2\Re\{(\negVeryTiny \mathbf{w}_{\text{u},k}^{[n]})^H \mathbf{h}_k\mathbf{h}_k^H \mathbf{w}_{\text{u},k}\}}{\beta_{\text{u},k}^{[n]}} \negVeryTiny  \negVeryTiny - \negVeryTiny \negVeryTiny   \frac{|\mathbf{h}_k^H\mathbf{w}_{\text{u},k}^{[n]}|^2 \beta_{\text{u},k}}{(\beta_{\text{u},k}^{[n]})^2} \negVeryTiny \negVeryTiny \geq \negVeryTiny \negVeryTiny \rho_{\text{u},k}, \negVeryTiny \, \forall \, \negVeryTiny \subalign{&k \in \set{K},\\& u \in \{c,p\}}, \label{eq:cons_betaRhoCP_linearized}
\end{equation}
where $\mathbf{w}_{\text{u},k}^{[n]}$ and $\beta_{\text{u},k}^{[n]}$ denote the expansion point of the Taylor series. The remaining non-convex constraints \eqref{eq:nonconvex_rhoEavesdrop} and \eqref{eq:nonconvex_eavesdropSINR} are addressed analogously, resulting in the linearized constraints:
\vspace{-1em}

\begin{subequations} \small
    \begin{align}
    & \label{eq:lineraized_betaRhoCPfrac}  1 + \rho_{k',k} \leq 2^{\alpha_{k',k}^{[n]}} [ 1+ \ln(2) (\alpha_{k',k}-\alpha_{k',k}^{[n]})], \, \forall \; \subalign{&k \in \set{K}\\&k' \in \set{K} \setminus \{k\}},  \\[4pt]
    & 2 \Re \left\{ [\mathbf{w}_{\text{p},k}^{[n]}]^H \mathbf{h}_{k} \mathbf{h}_{k}^H  \mathbf{w}_{\text{p},k} \right\} - [\mathbf{w}_{\text{p},k}^{[n]}]^H \mathbf{h}_{k} \mathbf{h}_{k}^H  \mathbf{w}_{\text{p},k}^{[n]}, \nonumber \\[\extraYSpaceBetweenEquations]
    & + 2 \rho_{k',k}^{[n]} \left(\vphantom{\sum}\right. \negSpaceFormula \! \sum_{\raisebox{\extraDistSum}{$\scriptstyle j \in \mathcal{K} \setminus \{k', k\}$}}  \negSpaceFormula \negVeryTiny \left( [\mathbf{w}_{\text{p},j}^{[n]}]^H \mathbf{h}_{k} \mathbf{h}_{k}^H \mathbf{w}_{\text{p},j}^{[n]} \negVeryTiny - \negVeryTiny \Re \left\{[\mathbf{w}_{\text{p},j}^{[n]}]^H \mathbf{h}_{k} \mathbf{h}_{k}^H  \mathbf{w}_{\text{p},j} \right\} \right) \!\! \negVeryTiny \left.\vphantom{\sum}\right) \nonumber \\
    &  \label{eq:cons_secRate_linearised}  - \rho_{k',k} \! \left(\vphantom{\sum}\right. \negSpaceFormula \! \negVeryTiny \sum_{\raisebox{\extraDistSum}{$\scriptstyle j \in \mathcal{K} \setminus \{k', k\}$}} \negSpaceFormula \negVeryTiny  \left( [\mathbf{w}_{\text{p},j}^{[n]}]^H \mathbf{h}_{k} \mathbf{h}_{k}^H \mathbf{w}_{\text{p},j}^{[n]} \right) \negVeryTiny + \negVeryTiny \sigma^2_n \! \negVeryTiny \left.\vphantom{\sum}\right) \! \leq 0 , \negVeryTiny \forall \; \subalign{&k \in \set{K},\\&k' \in \set{K} \setminus \{k\}},       
    \end{align}
\end{subequations}
where $\mathbf{w}_{\text{p},k}^{[n]}$, $\alpha_{k',k}^{[n]}$ and $\rho_{k',k}^{[n]}$ denote the expansion points of the respective Taylor series. Finally, we define the reformulated convex optimization problem $\mathrm{P2}$ as: \looseness=-1
\begin{subequations} 
\label{eq:P2Final}
    \begin{align}
        & \mathrm{P2}\!: \max_{\mathbf{c},\mathbf{W}, \bm{\alpha}, \bm{\beta}, \bm{\rho}}   \,   \sum_{k \in \set{K}} \left( C_k + \alpha_{\text{p},k} - R_{\text{i},k}^{\mathrm{req}}\nonumber \right) \\
        & \hspace{\subjectToSpaceOptprob} \textrm{s.t.}  \, \eqref{eq:constC_TXPwr}, \eqref{eq:constC_cNoneg} , \eqref{eq:alphaSecrecyRateDiff}-\eqref{eq:convex_rhoEavesdrop},\eqref{eq:cons_betaCSum}, \eqref{eq:cons_betaPSum},\eqref{eq:cons_betaRhoCP_linearized}, \eqref{eq:lineraized_betaRhoCPfrac},\eqref{eq:cons_secRate_linearised} \nonumber
    \end{align}
\end{subequations}

%% file: chapters/04_Iterative_Algorithm.tex
\section{Iterative Algorithm}
\label{sec:iteralgo}
As a result of the linearization, the solution obtained from $\mathrm{P2}$ is no longer guaranteed to be globally or even locally optimal, as its optimality now strongly depends on the chosen expansion point tuple $\bm{\nu}^{[n]} = [\bm{\alpha}^{[n]},\, \mathbf{W}^{[n]},\, \bm{\beta}^{[n]},\, \bm{\rho}^{[n]}]$. This issue can be addressed by utilizing Algorithm~\ref{alg:inner_opt}, which converges to a locally optimal solution by solving consecutive instances of $\mathrm{P2}$.
Each instance $n$ of $\mathrm{P2}$ utilizes the expansion point tuple $\bm{\nu}^{[n]}$, which is obtained from the previous optimal expansion point tuple $\bm{\nu}^{*}$. The algorithm terminates when the \ac{SE} improvement between consecutive iterations falls below the threshold $\tau$. At this point, stationarity has been reached and the optimal $\text{SE}^{*}$, the optimal common rate allocation vector $\ma{c}^*$ and the optimal expansion point tuple $\bm{\nu}^{*}$ are returned.\looseness=-1\\
\textit{Convergence:} The convergence argument of \cite{Xia2022} applies to Algorithm~\ref{alg:inner_opt} and guarantees convergence to a locally optimal point, provided the initial problem ($n=0$) is feasible.\looseness=-1\\
\textit{Initialization:} The initial expansion point tuple $\bm{\nu}^{[0]} = (\bm{\alpha}^{[0]}$, $\mathbf{W}^{[0]}$, $\bm{\beta}^{[0]}$, $\bm{\rho}^{[0]})$ is chosen according to \cite{Xia2022}. The private stream precoder of each user $k$ is initialized as $\ma{w}_{\text{p},k}^{[0]} = q_{\text{p}} \frac{\ma{h}_k}{||\ma{h}_k||}$, with $q_{\text{p}}  = \eta \cdot \sqrt{\lambda P_{\text{t}} \cdot K^{-1}}$ and $0\leq \,\{\lambda,\eta\} \leq 1$. The common stream precoder $\ma{w}_{\text{c}}$ is initialized as $\ma{w}_{\text{c}}^{[0]} = q_{\text{c}} \ma{u}_{\text{c}}$, where $q_{\text{c}}  = \eta \cdot \sqrt{(1-\lambda) P_{\text{t}}}$ and $\ma{u}_{\text{c}}$ is the largest left singular vector of the channel matrix $\ma{H} = [\ma{h}_1,\ldots,\ma{h}_K]$. The variable $\lambda$ is used to allocate power between the initial common and private precoders, while $\eta$ scales the initial power within $[0,P_{\text{t}}]$. The variables $\bm{\rho}^{[0]}$ and $\bm{\beta}^{[0]}$ are initialized by replacing the inequalities \eqref{eq:nonconvex_eavesdropSINR}, \eqref{eq:cons_betaCSum} and \eqref{eq:cons_betaPSum} with equalities. The initial ${\bm{\alpha}}^{[0]}$ is then constructed from $\bm{\rho}^{[0]}$ by assuming equality in constraint \eqref{eq:nonconvex_rhoEavesdrop}. \looseness=-1
\vspace{-0.5em}
\let\oldAlCapFnt\AlCapFnt
\SetAlCapFnt{\fontsize{9.9}{11}\selectfont}
\begin{algorithm} \small
\begin{minipage}{\textwidth}
\caption{\fontsize{9.9}{11}\selectfont Iterative hybrid-RSMA precoder optimisation}
\end{minipage}
\label{alg:inner_opt}
\KwIn{$\tau$, \, $R_{\text{s},k}^{\mathrm{req}}$, \, $R_{\text{i},k}^{\mathrm{req}}$, \, $P_{\text{t}}$, \, $\bm{\nu}^{[0]}$ } 
$n \gets 0$; $\text{SE}^{*} \gets -\infty$\;
\Repeat{
$|\text{SE}^{*} - \text{SE}^{[n-1]}| \leq \tau \;$
}{
$\text{SE}^{[n]} \gets \text{SE}^{*}$\;
$\text{SE}^{*}, \ma{c}^*, \bm{\nu}^{*} \gets \mathrm{P2}\left(\bm{\nu}^{[n]}\right)$ \Comment*[r]{Solve $\mathrm{P2}$}
$\bm{\nu}^{[n+1]} \gets \bm{\nu}^{*}$;  $n \gets n + 1$ \; 
}
\Return{$\text{SE}^{*}, \ma{c}^*, \bm{\nu}^{*}$}
\end{algorithm}
\SetAlCapFnt{\oldAlCapFnt}
\vspace{-1.2em}

%% file: chapters/05_Results.tex
\section{Performance Results}
\label{sec:perfRes}
This section evaluates and compares the \ac{SE} performance of the proposed hybrid-PLS \ac{RSMA} system (\textit{Rhyb}) with existing methods in the \ac{FSDT} scenario. Besides Rhyb, we also evaluate the partially secure \ac{RSMA} system (\textit{Rpar}) from~\cite{Xia2022}, and a secure \ac{SDMA} implementation (\textit{SSDMA}) subject to the secrecy constraint~\eqref{eq:constC_reqSecrate}. \ac{NOMA}-based systems are not considered, since \ac{NOMA} inherently compromises data security by requiring at least one user to decode a message intended for another user. In the remainder of this section, we first evaluate the performance analytically and derive upper and lower bounds on the SE gain. We then validate our results numerically using CVX in conjunction with the MOSEK solver.

\subsection{Analytical Results}
\label{sec:theoPerf}
Both Rhyb and SSDMA provide \ac{FSDT}, whereas Rpar only secures data in the private streams. Thus, the effective \ac{SE} of Rpar is given by $\text{SE}_{\text{Rpar}} = \sum_{k \in \set{K}} R_{\text{p},k}$, while the effective \ac{SE} of Rhyb is given by $\text{SE}_{\text{Rhyb}} = \sum_{k \in \set{K}} ( C_k + R_{\text{p},k} - R_{\text{i},k}^{\mathrm{req}} )$, as defined  in \eqref{eq:constC_objFunc}. Using these expressions, we define the \ac{SE} gain $\Delta_{\text{SE}}$ of Rhyb over Rpar as \looseness=-1
\begin{align}
    \Delta_{\text{SE}} & = \sum_{k \in \mathcal{K}} \left( C_k + R^{\text{Rhyb}}_{\text{p},k} - R_{\text{i},k}^{\mathrm{req}} \right) - \sum_{k \in \mathcal{K}} R^{\text{Rpar}}_{\text{p},k}, \nonumber \\
    & \geq \sum_{k \in \mathcal{K}} \left( C_k - R_{\text{i},k}^{\mathrm{req}} \right),
\end{align}
since $R^{\text{Rhyb}}_{\text{p},k} \geq R^{\text{Rpar}}_{\text{p},k}$ due to constraint~\eqref{eq:constC_reqBaseRate}. Since $\sum_{k \in \mathcal{K}} R_{\text{i},k}^{\mathrm{req}}$ is constant, $\Delta_{\text{SE}}$ is determined by the allocated common rate $R_c \! = \!\sum_{k \in \mathcal{K}} C_k$, leading to two operational scenarios. In the \ac{NOMA} region of operation~\cite{Mao2022}, the majority of power is allocated to the common stream, resulting in a significant SE gain $\Delta_{\text{SE}}$. The SE gain upper bound $\overline{\Delta}_{\text{SE}}$ is attained when RSMA reduces to NOMA, i.e., when all power is allocated to the common stream. In this scenario, Rpar yields $\text{SE}_{\text{Rpar}}  =  0$, while Rhyb achieves $\text{SE}_{\text{Rhyb}}  \approx \sum_{k \in \mathcal{K}} C_k$, as $\sum_{k \in \mathcal{K}} R_{\text{i},k}^{\mathrm{req}}$ is expected to be insignificant in practical scenarios. In the \ac{SDMA} regime~\cite{Mao2022} most power is allocated to the private streams, resulting in a low $R_c$ and minor \ac{SE} gains. The \ac{SE} gain lower bound $\underline{\Delta}_{\text{SE}}$ is attained when RSMA reduces to SDMA  and thus no power is allocated to the common stream ($R_c = 0$), yielding $\underline{\Delta}_{\text{SE}} = - \sum_{k \in \mathcal{K}} R^{\mathrm{req}}_{\text{i},k}$. However, since $R^{\mathrm{req}}_{\text{i},k} \, \forall  k \in \mathcal{K}$ is insignificant in practice, the SE loss is expected to be minimal.\looseness=-1

A similar trend is observed when comparing Rhyb with SSDMA. In the \ac{NOMA} region, Rhyb achieves notably higher \ac{SE} compared to SSDMA, which cannot exploit the common message. In the SDMA region, Rpar becomes equivalent to SSDMA, causing $\Delta_{\text{SE}}$ to approach the same lower bound $\underline{\Delta}_{\text{SE}}$.\looseness=-1

\subsection{Numerical Results}
\label{sec:numRes}
To verify the theoretical results, we examine the impact of the (i) channel conditions and (ii) secrecy rate requirements on the \ac{SE}, as well as (iii) the average \ac{SE} over random channel realizations. All experiments were conducted in a two-user scenario ($K=2$), with the \ac{BS} employing either two ($N=2$) or twelve ($N=12$) antennas. The system parameters were set to $\sigma_{\text{n}}^2 = 1$ and $P_{\text{t}} = 20$ dB, while the initialization variables were chosen as $\lambda = 0.5$ and $\eta = 0.01$. \looseness=-1

\newcommand{\subcapvspace}{\vspace{-1.5pt}}
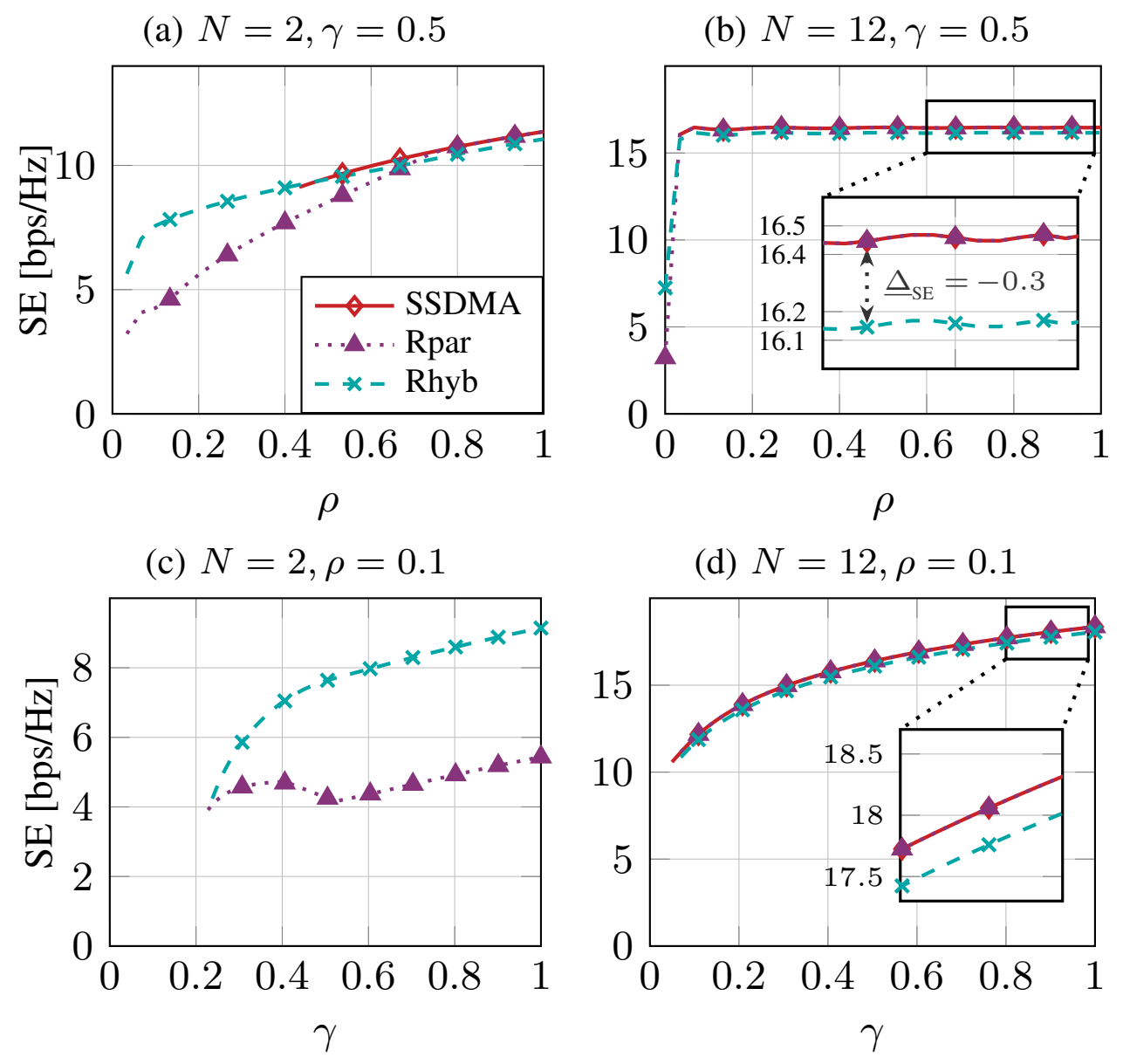
\begin{figure}[t]
  \vspace*{0.20in} 
  \centering
  \subfloat[\subcapvspace$N=2, \gamma = 0.5$]{%
      \input{plots/TC1_rho_under_res}\label{fig:TC1}}
  \hfil\hspace{-5pt}
  \subfloat[\subcapvspace $N=12, \gamma = 0.5$]{%
      \input{plots/TC2_rho_over_res}\label{fig:TC2}}
  \vspace{\plotdistRedVert}
  \subfloat[\subcapvspace $N=2, \rho = 0.1$]{%
      \input{plots/TC3_gamma_under_res}\label{fig:TC3}}
  \hfil\hspace{-9pt}
  \subfloat[\subcapvspace $N=12, \rho = 0.1$]{%
      \input{plots/TC4_gamma_over_res}\label{fig:TC4}}
  \vspace{-6pt}
  \caption{Channel influence for $R^{\mathrm{req}}_{\text{i}} = [0.2,0.1], R^{\mathrm{req}}_{\text{s}} = [2,1]$}
  \label{fig:All_TC_firstBlock}
\end{figure}

\textit{(i) Channel influence}: Fig.~\ref{fig:All_TC_firstBlock} illustrates the influence of the relative channel angle $\rho$ and the relative channel strength $\gamma$ on the \ac{SE} performance. Both parameters are defined according to the channel model from~\cite{Xia2022, Mao2022}: 
\begin{equation} \small
    \begin{aligned}
    &\mathbf{h}_1 = \begin{bmatrix} 1, 1, \ldots, 1 \end{bmatrix}^{H}, \\
    &\mathbf{h}_2 = \gamma \cdot \begin{bmatrix} 1, e^{j \theta}, \ldots, e^{j (N-1)\theta}\end{bmatrix}^{H},\\
    &\rho = 1- \frac{|\mathbf{h}_1^H \mathbf{h}_2|}{||\mathbf{h}_1||^2||\mathbf{h}_2||^2} \, \in \, [0,1],
    \end{aligned}
\end{equation}
where $\theta$ denotes the absolute channel angle. The influence of $\rho$ is examined in Fig. \ref{fig:TC1} and Fig. \ref{fig:TC2} for $\gamma = 0.5$, whereas the influence of $\gamma$ is examined in Fig. \ref{fig:TC3} and Fig. \ref{fig:TC4} for $\rho = 0.1$. Both evaluations assume a key information rate of $R^{\mathrm{req}}_{\text{i}} = [0.2,0.1]$ and a secrecy requirement of $R^{\mathrm{req}}_{\text{s}} = [2,1]$. \looseness=-1

The numerical results highlight that Rhyb achieves a significant \ac{SE} gain (${\Delta}_{\text{SE}}$) over Rpar, when the number of antennas per user is small ($N=2$), the channels are aligned (small $\rho$), and the channel gains are symmetric ($\gamma$ close to~1). This is consistent with the theoretical analysis, as these parameters correspond to the \ac{NOMA} region of operation~\cite{Mao2022}.
In contrast, ${\Delta}_{\text{SE}}$ decreases for larger numbers of antennas ($N=12$), orthogonal channels ($\rho$ close to 1), and asymmetric channel gains ($\gamma$ not close to 1). This is also consistent with earlier results, as these parameters correspond to the \ac{SDMA} 
region of operation~\cite{Mao2022}. As the SDMA characteristics become increasingly pronounced, ${\Delta}_{\text{SE}}$ decreases, until the lower bound $\underline{\Delta}_{\text{SE}} = -0.3$ is reached. Fig. \ref{fig:TC1} and Fig. \ref{fig:TC3} further show that SSDMA is unable to find feasible solutions in the \ac{NOMA} regime, as it cannot utilize a common message. \looseness=-1
\begin{figure}[t]
  \vspace*{0.20in}
  \centering
  \subfloat[\subcapvspace$N=2$, $\rho=0.1$, $\gamma = 0.5$]{%
      \input{plots/TC5_rsec_under_res}\label{fig:TC5}}
  \hfil\hspace{-15pt}
  \subfloat[\subcapvspace $N=12$, $\rho=0.1$, $\gamma = 0.5$]{%
      \input{plots/TC6_rsec_over_res}\label{fig:TC6}}
  \vspace{\plotdistRedVert-2pt}
  \subfloat[\subcapvspace $N=2$, $\rho=0.1$, $\gamma = 0.5$]{%
      \input{plots/TC7_indiv_under_res}\label{fig:TC7}}
  \hfil\hspace{-15pt}
  \subfloat[\subcapvspace $N=12$, $\rho=0.1$, $\gamma = 0.5$]{%
      \input{plots/TC8_indiv_over_res}\label{fig:TC8}}
  \vspace{-6pt}
  \caption{Secrecy rate influence for $R^{\mathrm{req}}_{\text{i}} = [0.2,0.1]$, $R^{\mathrm{req}}_{\text{s},2} = 0.5 \cdot R^{\mathrm{req}}_{\text{s},1}$}
  \label{fig:All_TC_secondBlock}
\end{figure}
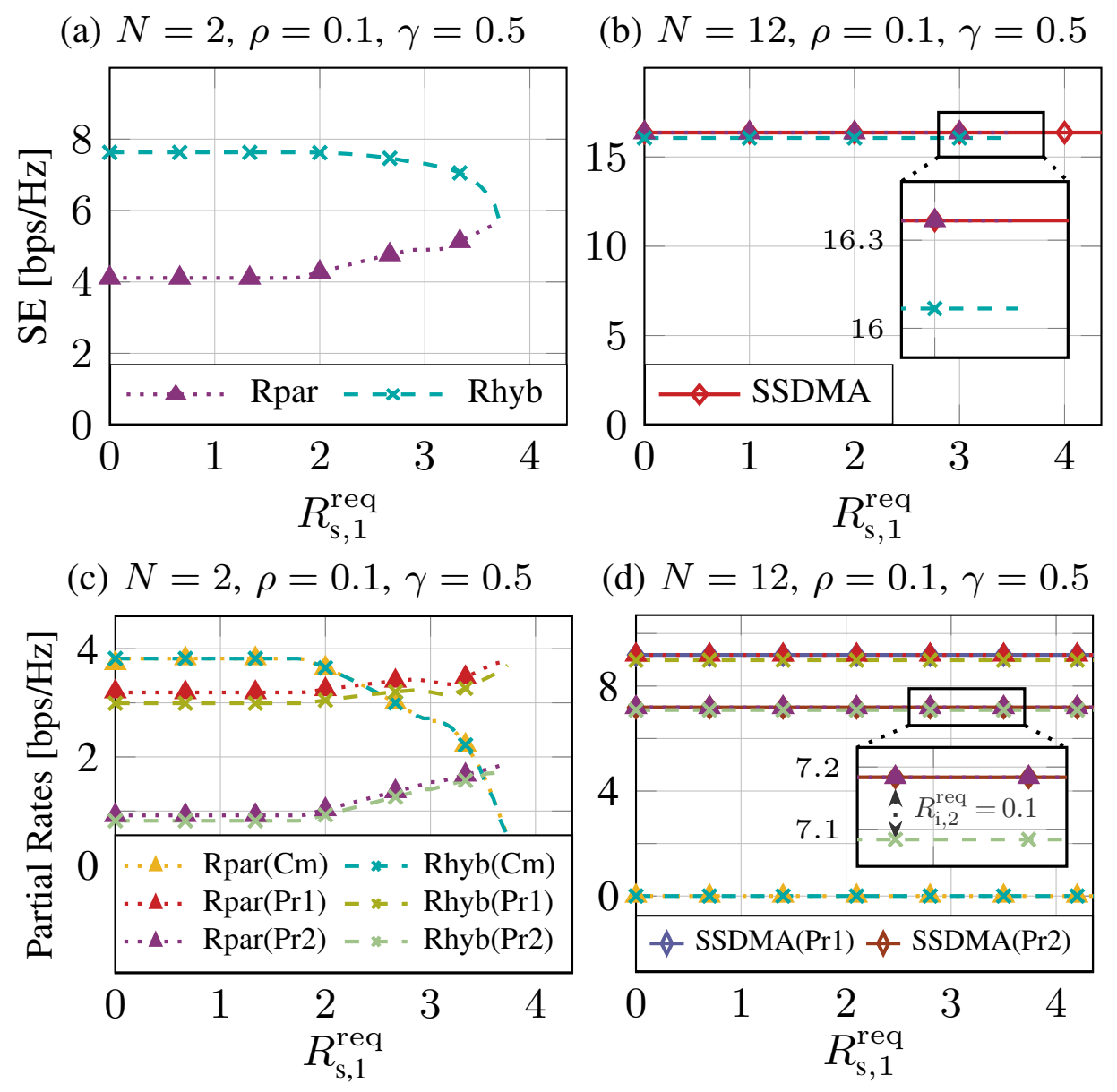

\textit{(ii) Secrecy Rates}: Fig. \ref{fig:All_TC_secondBlock} depicts the influence of the secrecy requirements on the SE performance (Fig. \ref{fig:TC5} and \ref{fig:TC6}) and on  the individual rate distributions (Fig. \ref{fig:TC7} and \ref{fig:TC8}). All simulations were conducted for $\rho = 0.1$, $\gamma = 0.5$, asymmetric key information $R^{\mathrm{req}}_{\text{i}} = [0.2, 0.1]$ and secrecy requirements $R^{\mathrm{req}}_{\text{s},2} = 0.5 \cdot R^{\mathrm{req}}_{\text{s},1}$ to reflect the asymmetric channel gains. 
The key information rate $R^{\mathrm{req}}_{\text{i},k}$ of each user $k$ was subtracted from the respective private rate of Rhyb in Fig. \ref{fig:TC7} and \ref{fig:TC8}. \looseness=-1

The results demonstrate that for $N=2$ and low $R^{\mathrm{req}}_{\text{s},1}$ a significant ${\Delta}_{\text{SE}}$ is achieved. This is attributed to the channel configuration, which belongs to the \ac{NOMA} region, as indicated by the significant common rate observed in Fig. \ref{fig:TC7}. Increasing $R^{\mathrm{req}}_{\text{s},1}$ requires allocating more power to the private streams, which in turn leads to a transition to the \ac{SDMA} regime and a reduction in ${\Delta}_{\text{SE}}$. Increasing $R^{\mathrm{req}}_{\text{s},1}$ in a scenario with $N=12$ antennas, does not affect the power allocation or ${\Delta}_{\text{SE}}$ since all power is already allocated to the private streams, as illustrated in Fig. \ref{fig:TC8}. \looseness=-1
Additionally, Fig. \ref{fig:TC6} indicates that Rhyb and Rpar are unable to find feasible solutions for $R^{\mathrm{req}}_{\text{s}} \geq [3.6, 1.8]$, even though SSDMA demonstrates their existence. 
Since SSDMA is a subscheme of Rpar, all SSDMA-feasible solutions are also feasible for Rpar when $R_c\!=\!0$. Therefore, the discrepancy is most likely attributable to the increased complexity of the Rhyb and Rpar optimization problems relative to SSDMA. \looseness=-1

\begin{figure}[!t]
\vspace*{-0.1in}
\centering
\subfloat[\subcapvspace$N=2$]{%
    \input{plots/TC9_Rsec_rnd_under}
    \label{fig:TC9}}
\hfil
\hspace{-5pt}
\subfloat[\subcapvspace $N=12$]{%
    \input{plots/TC10_Rsec_rnd_over}
    \label{fig:TC10}}
\vspace{\plotdistRedVert}
\subfloat[\subcapvspace $N=2$]{%
    \input{plots/TC11_Rsec_rnd_over_infeas}
    \label{fig:TC11}}
\hfil
\hspace{-9pt}
\subfloat[\subcapvspace $N=12$]{%
    \input{plots/TC12_Rsec_rnd_over_infeas}
    \label{fig:TC12}}
\caption{Random channel evaluation for $R^{\mathrm{req}}_{\text{i}} \! = \! [0.2,0.2]$, $R^{\mathrm{req}}_{\text{s},2} \! = \! R^{\mathrm{req}}_{\text{s},1}$}
\label{fig:All_TC_thirdBlock}
\end{figure}
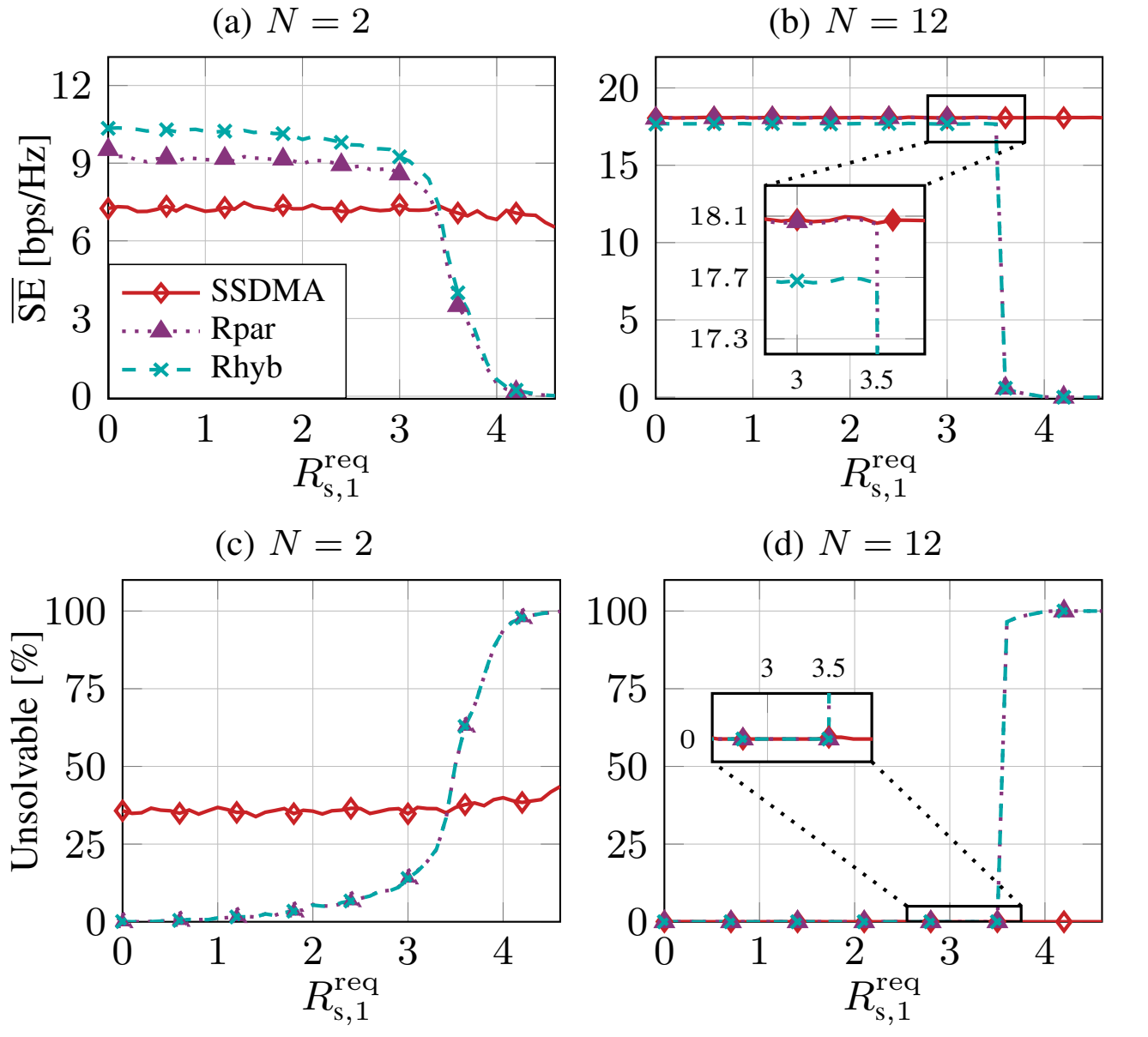

\textit{(iii) Random Channels}: Fig. \ref{fig:All_TC_thirdBlock} presents the evaluation of the methods under random channel realizations, with Fig. \ref{fig:TC9} and Fig. \ref{fig:TC10} depicting the average \ac{SE} performance. The remaining figures, Fig. \ref{fig:TC11} and Fig. \ref{fig:TC12}, illustrate the percentage of channel realizations for which no feasible solution could be found. The average \ac{SE} ($\overline{\text{SE}}$) was calculated over 2500 channels where an \ac{SE} of zero was assigned if no feasible solution could be found for a channel realization. In all cases, symmetric secrecy requirements ($R^{\text{req}}_{\text{s},1} = R^{\text{req}}_{\text{s},2}$) and symmetric key information requirements ($R^{\text{req}}_{\text{i}} = [0.2, 0.2]$) were assumed. \looseness=-1

Similar to previous findings, we observe a significant SE gain ($\negVeryTiny{\Delta}_{\text{SE}} \!   \approx    \! 1\negVeryTiny$) for a low number of antennas ($\negVeryTiny \negVeryTiny N\!  = \!  2$) and small-to-medium secrecy requirements. Although increasing $R^{\text{req}}_{\text{s},1}$ transitions to the \ac{SDMA} regime,  ${\Delta}_{\text{SE}}$ only marginally decreases until $R^{\text{req}}_{\text{s},1} \! \geq \negVeryTiny \negVeryTiny  3.5$. These results demonstrate that Rhyb provides a consistent performance increase, with gains of up to 13\%, independent of channel realizations, provided that other parameters (e.g., $N$) place the system in the overall \ac{NOMA} regime. 
For $N\!=\!12$, $\Delta_{\text{SE}}$ remains at its lower bound, $\underline{\Delta}_{\text{SE}}\negVeryTiny=\negVeryTiny-0.4$ for $R^{\text{req}}_{\text{s},1} \leq 3.5$, corresponding to a 2\% performance loss. We further observe in Fig. \ref{fig:TC11} and \ref{fig:TC12} that for $R^{\text{req}}_{\text{s},1} \negVeryTiny \geq \negVeryTiny 3.5$, the percentage of infeasible channel realizations increases steeply. This likely results from the complexity of the Rhyb and Rpar optimization problems, as discussed in (ii). Since unsolvable problems are recorded with an \ac{SE} of zero, this also leads to a steep decline of the \ac{SE} performance, as shown in Fig. \ref{fig:TC9} and \ref{fig:TC10}. \looseness=-1

%% file: plots/TC1_rho_under_res.tex
\begin{tikzpicture}

    \def\legendLinewidth{0.8pt}
    \def\legendMarkSize{1.9pt}
    
	\begin{axis}[
		width=138, 
		height=120,
		xmin = 0, xmax = 1,
		ymin = 0, ymax = 14,
		xlabel={$\rho$},
		ylabel={SE [bps/Hz]},
		grid=both,
		grid style={line width=.1pt, draw=gray!10},
		major grid style={line width=.2pt,draw=gray!50},
        legend style={
          font=\fontsize{7.3}{7.3}\selectfont,
          at={(1,0)},
          anchor=south east,
          legend columns=1,
          cells={anchor=west},
          inner xsep=2.5pt,
          inner ysep=1pt,
          row sep=-2pt,
          scale=1,
        },
		ymode=linear,
		xlabel style={font=\fontsize{\fontsizevar}{\fontsizevar}\selectfont},
		ylabel style={font=\fontsize{\fontsizevarsYLabel}{\fontsizevarsYLabel}\selectfont},
		tick label style={font=\fontsize{\fontsizevar}{\fontsizevar}\selectfont},
		xlabel style={at={(axis description cs:0.5,-0.15)},rotate=0,anchor=south},
		ylabel style={at={(axis description cs:0.28,0.5)},rotate=0,anchor=south}
		]

        \addplot [
            color=RMSEColour3,
            line width=\plotLinewidth,
            no markers,
            forget plot
        ]
            table [
                x=x_val,
                y=SDMA_WSR_y,
                col sep=comma,
            ]{plots/rawData/TC3RhoUnderCleaned.csv};
        
        \addplot [
            only marks,
            mark=diamond,
            color=RMSEColour3,
            mark size=\plotMarkersize,
            line width=\plotLinewidth, 
            forget plot
        ]
            table [
                x=x_val,
                y=SDMA_WSR_y,
                col sep=comma,
            ]{plots/markers/TC3Markers.csv};

        \addplot [
            color=RMSEColour2,
            dotted,
            line width=\plotLinewidth,
            no markers,
            forget plot
        ]
            table [
                x=x_val,
                y=RSMA_sec_SOA_PRECODER_ONLY_PRIVATE_RATES_COUNTED_WSR_y,
                col sep=comma,
            ]{plots/rawData/TC3RhoUnderCleaned.csv};
        
        \addplot [
            only marks,
            mark=triangle*,
            color=RMSEColour2,
            mark options={solid},
            mark size=\plotMarkersize,
            line width=\plotLinewidth, 
            forget plot
        ]
            table [
                x=x_val,
                y=RSMA_sec_SOA_PRECODER_ONLY_PRIVATE_RATES_COUNTED_WSR_y,
                col sep=comma,
            ]{plots/markers/TC3Markers.csv};

        \addplot [
            color=RMSEColour8,
            dashed,
            line width=\plotLinewidth,
            no markers,
            forget plot
        ]
            table [
                x=x_val,
                y=RSMAhyb_WSR_y,
                col sep=comma,
            ]{plots/rawData/TC3RhoUnderCleaned.csv};
        
        \addplot [
            only marks,
            mark=x,
            color=RMSEColour8,
            mark options={solid},
            mark size=\plotMarkersize,
            line width=\plotLinewidth,
            forget plot
        ]
            table [
                x=x_val,
                y=RSMAhyb_WSR_y,
                col sep=comma,
            ]{plots/markers/TC3Markers.csv};

        \addlegendimage{color=RMSEColour3,mark=diamond,mark options={solid, xscale=1.1, yscale=1}, line width=\legendLinewidth,mark size=\legendMarkSize}
        \addlegendentry{SSDMA}
    
        \addlegendimage{color=RMSEColour2,dotted,mark=triangle*,mark options={solid},line width=\legendLinewidth,mark size=\legendMarkSize}
        \addlegendentry{Rpar}
    
        \addlegendimage{color=RMSEColour8,dashed,mark=x,mark options={solid},line width=\legendLinewidth,mark size=\legendMarkSize}
        \addlegendentry{Rhyb}
		
	\end{axis}
\end{tikzpicture}

%% file: plots/TC2_rho_over_res.tex
\begin{tikzpicture}
    \def\markerxthirdlast{0.6666999999999998483}
    \def\markerxsecondlast{0.80001999999999995339}
    \def\markerxlast{0.93333999999999994746}

    \def\arrowyStart{16.458821870170652346}
    \def\arrowyEnd{16.158815450094735411}
    
    \def\zoomboxXmin{0.6}
    \def\zoomboxXmax{0.985}
    \def\zoomboxYmin{15}
    \def\zoomboxYmax{18}

    \pgfmathsetmacro{\zoomboxXmid}{0.5*(\zoomboxXmin+\zoomboxXmax)}
    \pgfmathsetmacro{\offsetmultiply}{0.5*(\zoomboxXmax-\zoomboxXmin)}
    \pgfmathsetmacro{\zoomboxlowermidleftx}{\zoomboxXmid - \offsetmultiply}
    \pgfmathsetmacro{\zoomboxlowermidrightx}{\zoomboxXmid + \offsetmultiply}

    \newcommand{\zoomboxLineWidth}{semithick}
    
    \def\zoomboxBorderColor{black}
    \def\zoomboxBgColor{none}
    \def\zoomboxOpacity{1}
    \def\zoomlinecolor{black}

    \def\legendLinewidth{0.8pt}

    \pgfmathsetmacro{\relmarkerxthirdlast}{(\markerxthirdlast-\zoomboxXmin)/(\zoomboxXmax-\zoomboxXmin)}
    \pgfmathsetmacro{\relmarkerxsecondlast}{(\markerxsecondlast-\zoomboxXmin)/(\zoomboxXmax-\zoomboxXmin)}
    \pgfmathsetmacro{\relmarkerxlast}{(\markerxlast-\zoomboxXmin)/(\zoomboxXmax-\zoomboxXmin)}

	\begin{axis}[
        name=mainplot,
		width=139, 
		height=120,
		xmin = 0, xmax = 1,
		ymin = 0, ymax = 20,
		xlabel={$\rho$},
		grid=both,
		grid style={line width=.1pt, draw=gray!10},
		major grid style={line width=.2pt,draw=gray!50},
		ymode=linear,
        ytick={0,5,10,15},
		xlabel style={font=\fontsize{\fontsizevar}{\fontsizevar}\selectfont},
		ylabel style={font=\fontsize{\fontsizevar}{\fontsizevar}\selectfont},
		tick label style={font=\fontsize{\fontsizevar}{\fontsizevar}\selectfont},
		xlabel style={at={(axis description cs:0.5,-0.15)},rotate=0,anchor=south},
		ylabel style={at={(axis description cs:0.31,0.5)},rotate=0,anchor=south}
		]

        \addplot [
            color=RMSEColour3,
            line width=\plotLinewidth,
            no markers,
            forget plot
        ]
            table [
                x=x_val,
                y=SDMA_WSR_y,
                col sep=comma,
            ]{plots/rawData/TC4RhoOverCleaned.csv};
        
        \addplot [
            only marks,
            mark=diamond,
            color=RMSEColour3,
            mark size=\plotMarkersize,
            line width=\plotLinewidth,
            forget plot
        ]
            table [
                x=x_val,
                y=SDMA_WSR_y,
                col sep=comma,
            ]{plots/markers/TC4Markers.csv};

        \addplot [
            color=RMSEColour2,
            dotted,
            line width=\plotLinewidth,
            no markers,
            forget plot
        ]
            table [
                x=x_val,
                y=RSMA_sec_SOA_PRECODER_ONLY_PRIVATE_RATES_COUNTED_WSR_y,
                col sep=comma,
            ]{plots/rawData/TC4RhoOverCleaned.csv};
        
        \addplot [
            only marks,
            mark=triangle*,
            color=RMSEColour2,
            mark options={solid},
            mark size=\plotMarkersize,
            line width=\plotLinewidth,
            forget plot
        ]
            table [
                x=x_val,
                y=RSMA_sec_SOA_PRECODER_ONLY_PRIVATE_RATES_COUNTED_WSR_y,
                col sep=comma,
            ]{plots/markers/TC4Markers.csv};

        \addplot [
            color=RMSEColour8,
            dashed,
            line width=\plotLinewidth,
            no markers,
            forget plot
        ]
            table [
                x=x_val,
                y=RSMAhyb_WSR_y,
                col sep=comma,
            ]{plots/rawData/TC4RhoOverCleaned.csv};
        
        \addplot [
            only marks,
            mark=x,
            color=RMSEColour8,
            mark options={solid},
            mark size=\plotMarkersize,
            line width=\plotLinewidth,
            forget plot
        ]
            table [
                x=x_val,
                y=RSMAhyb_WSR_y,
                col sep=comma,
            ]{plots/markers/TC4Markers.csv};

        \addplot [
            draw=\zoomboxBorderColor,
            fill=\zoomboxBgColor,
            opacity=\zoomboxOpacity,
            \zoomboxLineWidth
        ] 
        coordinates {
            (\zoomboxXmin,\zoomboxYmin)
            (\zoomboxXmax,\zoomboxYmin)
            (\zoomboxXmax,\zoomboxYmax)
            (\zoomboxXmin,\zoomboxYmax)
            (\zoomboxXmin,\zoomboxYmin)
        };
        
    \coordinate (zoombox_lower_mid_left) at (axis cs:\zoomboxlowermidleftx,\zoomboxYmin);
    \coordinate (zoombox_lower_mid_right) at (axis cs:\zoomboxlowermidrightx,\zoomboxYmin);
	\end{axis}

    \begin{axis}[
        name=subplot,
        at={(mainplot.south east)},
        axis lines=box,
        width=100, 
		height=82,
        xshift=-5pt,
        yshift=46.7pt,
		grid style={line width=.1pt, draw=gray!10},
		major grid style={line width=.2pt,draw=gray!50},
        anchor=north east,
        grid=both,
        xtick={0.8},
        xticklabels={},
        ytick={16.1, 16.2, 16.4, 16.5},
        yticklabels={16.1, 16.2, 16.4, 16.5},
        xmin=0.6, xmax=0.985,
        ymin=16, ymax=16.6,
        xlabel style={font=\fontsize{\tiksizeVarInnerPlot}{\tiksizeVarInnerPlot}\selectfont},
		ylabel style={font=\fontsize{\tiksizeVarInnerPlot}{\tiksizeVarInnerPlot}\selectfont},
		tick label style={font=\fontsize{\tiksizeVarInnerPlot}{\tiksizeVarInnerPlot}\selectfont},
        axis background/.style={fill=white},
        axis line style={\zoomboxLineWidth, draw=\zoomboxBorderColor}
       ]

        \addplot [
            color=RMSEColour3,
            line width=\plotLinewidth,
            no markers,
            forget plot,
            restrict x to domain=0.6:1
        ]
            table [
                x=x_val,
                y=SDMA_WSR_y,
                col sep=comma,
            ]{plots/rawData/TC4RhoOverCleaned.csv};
        
        \addplot [
            only marks,
            mark=diamond,
            color=RMSEColour3,
            mark size=\plotMarkersize,
            line width=\plotLinewidth, 
            forget plot,
            restrict x to domain=0.6:1
        ]
            table [
                x=x_val,
                y=SDMA_WSR_y,
                col sep=comma,
            ]{plots/markers/TC4Markers.csv};

        \addplot [
            color=RMSEColour2,
            dotted,
            line width=\plotLinewidth,
            no markers,
            forget plot,
            restrict x to domain=0.6:1
        ]
            table [
                x=x_val,
                y=RSMA_sec_SOA_PRECODER_ONLY_PRIVATE_RATES_COUNTED_WSR_y,
                col sep=comma,
            ]{plots/rawData/TC4RhoOverCleaned.csv};
        
        \addplot [
            only marks,
            mark=triangle*,
            color=RMSEColour2,
            mark options={solid},
            mark size=\plotMarkersize,
            line width=\plotLinewidth, 
            forget plot
        ]
            table [
                x=x_val,
                y=RSMA_sec_SOA_PRECODER_ONLY_PRIVATE_RATES_COUNTED_WSR_y,
                col sep=comma,
            ]{plots/markers/TC4Markers.csv};

        \addplot [
            color=RMSEColour8,
            dashed,
            line width=\plotLinewidth,
            no markers,
            forget plot,
            restrict x to domain=0.6:1
        ]
            table [
                x=x_val,
                y=RSMAhyb_WSR_y,
                col sep=comma,
            ]{plots/rawData/TC4RhoOverCleaned.csv};
        
        \addplot [
            only marks,
            mark=x,
            color=RMSEColour8,
            mark options={solid},
            mark size=\plotMarkersize,
            line width=\plotLinewidth, 
            forget plot,
            restrict x to domain=0.6:1
        ]
            table [
                x=x_val,
                y=RSMAhyb_WSR_y,
                col sep=comma,
            ]{plots/markers/TC4Markers.csv};

        \coordinate (innerplot_upper_left) at (rel axis cs:0,1);
        \coordinate (innerplot_outer_corner) at (rel axis cs:1,1);
            
        \draw[{Stealth[scale=0.6]}-{Stealth[scale=0.6]}, dotted, color=black!80, thick, shorten <=2.3pt]
        (axis cs:\markerxthirdlast,\arrowyStart)
        -- (axis cs:\markerxthirdlast,\arrowyEnd)
        node[midway, right, font=\fontsize{5.75}{8.5}\selectfont, yshift = -1pt, xshift = 0pt] {  $\underline{\Delta}_{\text{SE}} \negVeryTiny =  \negVeryTiny \negVeryTiny -0.3$};

    \end{axis}
    
    \pgfmathsetmacro{\relmarkerxthirdlast}{(\markerxthirdlast-\zoomboxXmin)/(\zoomboxXmax-\zoomboxXmin)}
    \pgfmathsetmacro{\relmarkerxlast}{(\markerxlast-\zoomboxXmin)/(\zoomboxXmax-\zoomboxXmin)}

    \draw [
        thick,
        dotted,
        color=\zoomlinecolor
    ] (zoombox_lower_mid_left) -- (innerplot_upper_left);
    
    \draw [
        thick,
        dotted,
        color=\zoomlinecolor
    ] (zoombox_lower_mid_right) -- (innerplot_outer_corner);     
\end{tikzpicture}

%% file: plots/TC3_gamma_under_res.tex
\begin{tikzpicture}
	\begin{axis}[
		width=138, 
		height=120,
		xmin = 0, xmax = 1,
		ymin = 0, ymax = 10,
		xlabel={$\gamma$},
		ylabel={SE [bps/Hz]},
		grid=both,
		grid style={line width=.1pt, draw=gray!10},
		major grid style={line width=.2pt,draw=gray!50},
        legend style={
          font=\fontsize{7.1}{7.1}\selectfont,
          at={(0,0)},
          anchor=south west,
          legend columns=2,
          cells={anchor=west},
          inner xsep=2.5pt,
          inner ysep=1pt,
          row sep=-2pt,
          scale=0.5,
          mark size=1.2pt,
          line width=0.3pt
        },
		ymode=linear,
		xlabel style={font=\fontsize{\fontsizevar}{\fontsizevar}\selectfont},
		ylabel style={font=\fontsize{\fontsizevarsYLabel}{\fontsizevarsYLabel}\selectfont},
		tick label style={font=\fontsize{\fontsizevar}{\fontsizevar}\selectfont},
		xlabel style={at={(axis description cs:0.5,-0.15)},rotate=0,anchor=south},
		ylabel style={at={(axis description cs:0.3,0.5)},rotate=0,anchor=south},
        ymode=linear,
        ytick={0,2,4,6,8},
		]
        \def\markrepeatrate{10}

        \addplot [
            color=RMSEColour3,
            line width=\plotLinewidth,
            no markers,
            forget plot
        ]
            table [
                x=x_val,
                y=SDMA_WSR_y,
                col sep=comma,
            ]{plots/rawData/TC7GammaUnderCleaned.csv};
        
        \addplot [
            only marks,
            mark=diamond*,
            color=RMSEColour3,
            mark size=\plotMarkersize,
            line width=\plotLinewidth,
            forget plot
        ]
            table [
                x=x_val,
                y=SDMA_WSR_y,
                col sep=comma,
            ]{plots/markers/TC7Markers.csv};

        \addplot [
            color=RMSEColour2,
            dotted,
            line width=\plotLinewidth,
            no markers,
            forget plot
        ]
            table [
                x=x_val,
                y=RSMA_sec_SOA_PRECODER_ONLY_PRIVATE_RATES_COUNTED_WSR_y,
                col sep=comma,
            ]{plots/rawData/TC7GammaUnderCleaned.csv};
        
        \addplot [
            only marks,
            mark=triangle*,
            color=RMSEColour2,
            mark options={solid},
            mark size=\plotMarkersize,
            line width=\plotLinewidth,
            forget plot
        ]
            table [
                x=x_val,
                y=RSMA_sec_SOA_PRECODER_ONLY_PRIVATE_RATES_COUNTED_WSR_y,
                col sep=comma,
            ]{plots/markers/TC7Markers.csv};

        \addplot [
            color=RMSEColour8,
            dashed,
            line width=\plotLinewidth,
            no markers,
            forget plot
        ]
            table [
                x=x_val,
                y=RSMAhyb_WSR_y,
                col sep=comma,
            ]{plots/rawData/TC7GammaUnderCleaned.csv};
        
        \addplot [
            only marks,
            mark=x,
            color=RMSEColour8,
            mark options={solid},
            mark size=\plotMarkersize,
            line width=\plotLinewidth,
            forget plot
        ]
            table [
                x=x_val,
                y=RSMAhyb_WSR_y,
                col sep=comma,
            ]{plots/markers/TC7Markers.csv};

  	\end{axis}          

\end{tikzpicture}

%% file: plots/TC4_gamma_over_res.tex
\begin{tikzpicture}

    \def\markerxsecondlast{0.80001999999999995339}
    \def\markerxlast{0.93333999999999994746}

    \def\zoomboxXmin{0.8}
    \def\zoomboxXmax{0.985}
    \def\zoomboxYmin{16.5}
    \def\zoomboxYmax{19.5}

    \pgfmathsetmacro{\zoomboxXmid}{0.5*(\zoomboxXmin+\zoomboxXmax)}
    \pgfmathsetmacro{\offsetmultiply}{0.5*(\zoomboxXmax-\zoomboxXmin)}
    \pgfmathsetmacro{\zoomboxlowermidleftx}{\zoomboxXmid - \offsetmultiply}
    \pgfmathsetmacro{\zoomboxlowermidrightx}{\zoomboxXmid + \offsetmultiply}
    
    \newcommand{\zoomboxLineWidth}{semithick}
    
    \def\zoomboxBorderColor{black}
    \def\zoomboxBgColor{none}
    \def\zoomboxOpacity{1}
    \def\zoomlinecolor{black}

    \def\legendLinewidth{0.8pt}

	\begin{axis}[
		width=141, 
		height=120,
		xmin = 0, xmax = 1,
		ymin = 0, ymax = 20,
		xlabel={$\gamma$},
		grid=both,
		grid style={line width=.1pt, draw=gray!10},
		major grid style={line width=.2pt,draw=gray!50},
		ymode=linear,
		xlabel style={font=\fontsize{\fontsizevar}{\fontsizevar}\selectfont},
		ylabel style={font=\fontsize{\fontsizevar}{\fontsizevar}\selectfont},
		tick label style={font=\fontsize{\fontsizevar}{\fontsizevar}\selectfont},
		xlabel style={at={(axis description cs:0.5,-0.15)},rotate=0,anchor=south},
		ylabel style={at={(axis description cs:0.31,0.5)},rotate=0,anchor=south},
        ymode=linear,
        ytick={0,5,10,15},
		]

        \addplot [
            color=RMSEColour3,
            line width=\plotLinewidth,
            no markers,
            forget plot
        ]
            table [
                x=x_val,
                y=SDMA_WSR_y,
                col sep=comma,
            ]{plots/rawData/TC8GammaOverCleaned.csv};
        
        \addplot [
            only marks,
            mark=diamond,
            color=RMSEColour3,
            mark size=\plotMarkersize,
            line width=\plotLinewidth,
            forget plot
        ]
            table [
                x=x_val,
                y=SDMA_WSR_y,
                col sep=comma,
            ]{plots/markers/TC8Markers.csv};

        \addplot [
            color=RMSEColour2,
            dotted,
            line width=\plotLinewidth,
            no markers,
            forget plot
        ]
            table [
                x=x_val,
                y=RSMA_sec_SOA_PRECODER_ONLY_PRIVATE_RATES_COUNTED_WSR_y,
                col sep=comma,
            ]{plots/rawData/TC8GammaOverCleaned.csv};
        
        \addplot [
            only marks,
            mark=triangle*,
            color=RMSEColour2,
            mark options={solid},
            mark size=\plotMarkersize,
            line width=\plotLinewidth,
            forget plot
        ]
            table [
                x=x_val,
                y=RSMA_sec_SOA_PRECODER_ONLY_PRIVATE_RATES_COUNTED_WSR_y,
                col sep=comma,
            ]{plots/markers/TC8Markers.csv};

        \addplot [
            color=RMSEColour8,
            dashed,
            line width=\plotLinewidth,
            no markers,
            forget plot
        ]
            table [
                x=x_val,
                y=RSMAhyb_WSR_y,
                col sep=comma,
            ]{plots/rawData/TC8GammaOverCleaned.csv};
        
        \addplot [
            only marks,
            mark=x,
            color=RMSEColour8,
            mark options={solid},
            mark size=\plotMarkersize,
            line width=\plotLinewidth,
            forget plot
        ]
            table [
                x=x_val,
                y=RSMAhyb_WSR_y,
                col sep=comma,
            ]{plots/markers/TC8Markers.csv};

        \addplot [
            draw=\zoomboxBorderColor,
            fill=\zoomboxBgColor,
            opacity=\zoomboxOpacity,
            \zoomboxLineWidth
        ] 
        coordinates {
            (\zoomboxXmin,\zoomboxYmin)
            (\zoomboxXmax,\zoomboxYmin)
            (\zoomboxXmax,\zoomboxYmax)
            (\zoomboxXmin,\zoomboxYmax)
            (\zoomboxXmin,\zoomboxYmin)
        };
        
        \coordinate (zoombox_lower_mid_left) at (axis cs:\zoomboxlowermidleftx,\zoomboxYmin);
        \coordinate (zoombox_lower_mid_right) at (axis cs:\zoomboxlowermidrightx,\zoomboxYmin);
    \end{axis}

    \begin{axis}[
        name=subplot,
        at={(mainplot.south east)},
        axis lines=box,
        width=80, 
		height=82,
        xshift=-5pt,
        yshift=46.7pt,
		grid style={line width=.1pt, draw=gray!10},
		major grid style={line width=.2pt,draw=gray!50},
        anchor=north east,
        grid=both,
        xtick={0.8},
        xticklabels={},
        ytick={17.5, 18, 18.5},
        xmin=0.8, xmax=0.985,
        ymin=17.3, ymax=18.7,
        xlabel style={font=\fontsize{\tiksizeVarInnerPlot}{\tiksizeVarInnerPlot}\selectfont},
		ylabel style={font=\fontsize{\tiksizeVarInnerPlot}{\tiksizeVarInnerPlot}\selectfont},
		tick label style={font=\fontsize{\tiksizeVarInnerPlot}{\tiksizeVarInnerPlot}\selectfont},
        axis background/.style={fill=white},
        axis line style={\zoomboxLineWidth, draw=\zoomboxBorderColor}
       ]

        \addplot [
            color=RMSEColour3,
            line width=\plotLinewidth,
            no markers,
            forget plot,
            restrict x to domain=0.6:1
        ]
            table [
                x=x_val,
                y=SDMA_WSR_y,
                col sep=comma,
            ]{plots/rawData/TC8GammaOverCleaned.csv};
        
        \addplot [
            only marks,
            mark=diamond,
            color=RMSEColour3,
            mark size=\plotMarkersize,
            line width=\plotLinewidth,
            forget plot,
            restrict x to domain=0.6:1
        ]
            table [
                x=x_val,
                y=SDMA_WSR_y,
                col sep=comma,
            ]{plots/markers/TC8Markers.csv};

        \addplot [
            color=RMSEColour2,
            dotted,
            line width=\plotLinewidth,
            no markers,
            forget plot,
            restrict x to domain=0.6:1
        ]
            table [
                x=x_val,
                y=RSMA_sec_SOA_PRECODER_ONLY_PRIVATE_RATES_COUNTED_WSR_y,
                col sep=comma,
            ]{plots/rawData/TC8GammaOverCleaned.csv};
        
        \addplot [
            only marks,
            mark=triangle*,
            color=RMSEColour2,
            mark options={solid},
            mark size=\plotMarkersize,
            line width=\plotLinewidth,
            forget plot
        ]
            table [
                x=x_val,
                y=RSMA_sec_SOA_PRECODER_ONLY_PRIVATE_RATES_COUNTED_WSR_y,
                col sep=comma,
            ]{plots/markers/TC8Markers.csv};

        \addplot [
            color=RMSEColour8,
            dashed,
            line width=\plotLinewidth,
            no markers,
            forget plot,
            restrict x to domain=0.6:1
        ]
            table [
                x=x_val,
                y=RSMAhyb_WSR_y,
                col sep=comma,
            ]{plots/rawData/TC8GammaOverCleaned.csv};
        
        \addplot [
            only marks,
            mark=x,
            color=RMSEColour8,
            mark options={solid},
            mark size=\plotMarkersize,
            line width=\plotLinewidth,
            forget plot,
            restrict x to domain=0.6:1
        ]
            table [
                x=x_val,
                y=RSMAhyb_WSR_y,
                col sep=comma,
            ]{plots/markers/TC8Markers.csv};
        \coordinate (innerplot_upper_left) at (rel axis cs:0,1);
        \coordinate (innerplot_outer_corner) at (rel axis cs:1,1);
  
    \end{axis}

    \draw [
        thick,
        dotted,
        color=\zoomlinecolor
    ] (zoombox_lower_mid_left) -- (innerplot_upper_left);

    \draw [
        thick,
        dotted,
        color=\zoomlinecolor
    ] (zoombox_lower_mid_right) -- (innerplot_outer_corner);  

\end{tikzpicture}

%% file: plots/TC5_rsec_under_res.tex
\begin{tikzpicture}
    \def\markrepeatrate{10}
    \def\maskingpoint{0.5}
    \def\xmin{0}
    \def\xmax{4.35}
    \def\ymin{0}
    \def\ymax{10}

	\begin{axis}[
		width=141, 
		height=120,
		xmin = \xmin, xmax = \xmax,
		ymin = \ymin, ymax = \ymax,
		xlabel={$R_{\text{s},1}^{\mathrm{req}}$},
		ylabel={SE [bps/Hz]},
		grid=both,
		grid style={line width=.1pt, draw=gray!10},
		major grid style={line width=.2pt,draw=gray!50},
        legend style={
          font=\fontsize{7.3}{7.3}\selectfont,
          at={(0,0)},
          anchor=south west,
          legend columns=2,
          cells={anchor=west},
          inner xsep=3pt,
          inner ysep=1pt,
          row sep=-2pt,
          column sep=3pt,
          scale=1,
          mark size=1.5pt,
          line width=0.3pt
        },
        legend image post style={xscale=1.21}, 
		ymode=linear,
        ytick={0,2,4,6,8},
        xtick={0,1,2,3,4},
		xlabel style={font=\fontsize{\fontsizevar}{\fontsizevar}\selectfont},
		ylabel style={font=\fontsize{\fontsizevarsYLabel}{\fontsizevarsYLabel}\selectfont},
		tick label style={font=\fontsize{\fontsizevar}{\fontsizevar}\selectfont},
		xlabel style={at={(axis description cs:0.5,-0.18)},rotate=0,anchor=south},
		ylabel style={at={(axis description cs:0.28,0.5)},rotate=0,anchor=south}
		]

        \addplot [mark=triangle*,
        dotted, 
        color = RMSEColour2,
        line width=0.8pt,
        mark repeat={\markrepeatrate},
        mark options={solid}] table [
		x=x_val,
		y=RSMA_sec_SOA_PRECODER_ONLY_PRIVATE_RATES_COUNTED_WSR_y,
		col sep=comma,
		]{plots/rawData/TC1RsecUnderCleaned.csv};

        \addplot [ mark=x, 
        dashed, 
        color = RMSEColour8, 
        line width=0.8pt, 
        mark repeat={\markrepeatrate},
        mark options={solid}]  table [
		x=x_val,
		y=RSMAhyb_WSR_y,
		col sep=comma,
		]{plots/rawData/TC1RsecUnderCleaned.csv};

		\legend{%
            Rpar,%
            Rhyb,%
		}
	\end{axis}
\end{tikzpicture}

%% file: plots/TC6_rsec_over_res.tex
\begin{tikzpicture}
    \def\TCSixLegendLinewidth{0.8pt}
    \def\TCSixLegendMarkSize{1.9pt}

    \def\zoomboxXmin{2.8}
    \def\zoomboxXmax{3.8}
    \def\zoomboxYmin{15}
    \def\zoomboxYmax{17.5}

    \pgfmathsetmacro{\zoomboxXmid}{0.5*(\zoomboxXmin+\zoomboxXmax)}
    \pgfmathsetmacro{\offsetmultiply}{0.5*(\zoomboxXmax-\zoomboxXmin)}
    \pgfmathsetmacro{\zoomboxlowermidleftx}{\zoomboxXmid - \offsetmultiply}
    \pgfmathsetmacro{\zoomboxlowermidrightx}{\zoomboxXmid + \offsetmultiply}

    \newcommand{\zoomboxLineWidth}{semithick}
    
    \def\zoomboxBorderColor{black}
    \def\zoomboxBgColor{none}
    \def\zoomboxOpacity{1}
    \def\zoomlinecolor{black}

    \def\legendLinewidth{0.8pt}

    \def\xmin{0}
    \def\xmax{4.35}
    \def\ymin{0}
    \def\ymax{20}
	\begin{axis}[
		width=141, 
		height=120,
		xmin = \xmin, xmax = \xmax,
		ymin = \ymin, ymax = \ymax,
		xlabel={$R_{\text{s},1}^{\mathrm{req}}$},
		grid=both,
		grid style={line width=.1pt, draw=gray!10},
		major grid style={line width=.2pt,draw=gray!50},
        legend style={
          font=\fontsize{7.3}{7.3}\selectfont,
          at={(0,0)},
          anchor=south west,
          legend columns=2,
          cells={anchor=west},
          inner xsep=2.5pt,
          inner ysep=1pt,
          row sep=-2pt,
          scale=0.5,
          mark size=1.2pt,
          line width=0.3pt
        },
        ymode=linear,
        ytick={0,5,10,15},
        xtick={0,1,2,3,4},
		xlabel style={font=\fontsize{\fontsizevar}{\fontsizevar}\selectfont},
		ylabel style={font=\fontsize{\fontsizevar}{\fontsizevar}\selectfont},
		tick label style={font=\fontsize{\fontsizevar}{\fontsizevar}\selectfont},
		xlabel style={at={(axis description cs:0.5,-0.18)},rotate=0,anchor=south},
		ylabel style={at={(axis description cs:0.5,0.5)},rotate=0,anchor=south}
		]

        \addplot [
            color=RMSEColour3,
            line width=\plotLinewidth,
            no markers,
            forget plot,
        ]
            table [
                x=x_val,
                y=SDMA_WSR_y,
                col sep=comma,
            ]{plots/rawData/TC2RsecOverCleaned.csv};
        
        \addplot [
            only marks,
            mark=diamond,
            color=RMSEColour3,
            mark size=\plotMarkersize,
            line width=\plotLinewidth,
            forget plot,
        ]
            table [
                x=x_val,
                y=SDMA_WSR_y,
                col sep=comma,
            ]{plots/markers/TC2Markers.csv};

        \addplot [
            color=RMSEColour2,
            dotted,
            line width=\plotLinewidth,
            no markers,
            forget plot,
        ]
            table [
                x=x_val,
                y=RSMA_sec_SOA_PRECODER_ONLY_PRIVATE_RATES_COUNTED_WSR_y,
                col sep=comma,
            ]{plots/rawData/TC2RsecOverCleaned.csv};
        
        \addplot [
            only marks,
            mark=triangle*,
            color=RMSEColour2,
            mark options={solid},
            mark size=\plotMarkersize,
            line width=\plotLinewidth,
            forget plot
        ]
            table [
                x=x_val,
                y=RSMA_sec_SOA_PRECODER_ONLY_PRIVATE_RATES_COUNTED_WSR_y,
                col sep=comma,
            ]{plots/markers/TC2Markers.csv};

        \addplot [
            color=RMSEColour8,
            dashed,
            line width=\plotLinewidth,
            no markers,
            forget plot,
        ]
            table [
                x=x_val,
                y=RSMAhyb_WSR_y,
                col sep=comma,
            ]{plots/rawData/TC2RsecOverCleaned.csv};
        
        \addplot [
            only marks,
            mark=x,
            color=RMSEColour8,
            mark options={solid},
            mark size=\plotMarkersize,
            line width=\plotLinewidth,
            forget plot,
        ]
            table [
                x=x_val,
                y=RSMAhyb_WSR_y,
                col sep=comma,
            ]{plots/markers/TC2Markers.csv};

        \addlegendimage{color=RMSEColour3,mark=diamond,mark options={solid, xscale=1.1, yscale=1}, line width=\TCSixLegendLinewidth,mark size=\TCSixLegendMarkSize}
        \addlegendentry{SSDMA}

        \addplot [
            draw=\zoomboxBorderColor,
            fill=\zoomboxBgColor,
            opacity=\zoomboxOpacity, 
            \zoomboxLineWidth
        ] 
        coordinates {
            (\zoomboxXmin,\zoomboxYmin)
            (\zoomboxXmax,\zoomboxYmin)
            (\zoomboxXmax,\zoomboxYmax)
            (\zoomboxXmin,\zoomboxYmax)
            (\zoomboxXmin,\zoomboxYmin)
        };
    \coordinate (zoombox_lower_mid_left) at (axis cs:\zoomboxlowermidleftx,\zoomboxYmin);
    \coordinate (zoombox_lower_mid_right) at (axis cs:\zoomboxlowermidrightx,\zoomboxYmin);

    \end{axis}

    \begin{axis}[
        name=subplot,
        at={(mainplot.south east)},
        axis lines=box,
        width=80, 
		height=82,
        xshift=-5pt,
        yshift=51.1pt,
		grid style={line width=.1pt, draw=gray!10},
		major grid style={line width=.2pt,draw=gray!50},
        anchor=north east,
        grid=both,
        xtick={3},
        xticklabels={},
        ytick={15.7, 16, 16.3},
        xmin=2.8, xmax=3.8, 
        ymin=15.9, ymax=16.5,
        xlabel style={font=\fontsize{\tiksizeVarInnerPlot}{\tiksizeVarInnerPlot}\selectfont},
		ylabel style={font=\fontsize{\tiksizeVarInnerPlot}{\tiksizeVarInnerPlot}\selectfont},
		tick label style={font=\fontsize{\tiksizeVarInnerPlot}{\tiksizeVarInnerPlot}\selectfont},
        axis background/.style={fill=white},
        axis line style={\zoomboxLineWidth, draw=\zoomboxBorderColor}
       ]

        \addplot [
            color=RMSEColour3,
            line width=\plotLinewidth,
            no markers,
            forget plot,
        ]
            table [
                x=x_val,
                y=SDMA_WSR_y,
                col sep=comma,
            ]{plots/rawData/TC2RsecOverCleaned.csv};
        
        \addplot [
            only marks,
            mark=diamond,
            color=RMSEColour3,
            mark size=\plotMarkersize,
            line width=\plotLinewidth,
            forget plot,
        ]
            table [
                x=x_val,
                y=SDMA_WSR_y,
                col sep=comma,
            ]{plots/markers/TC2Markers.csv};

        \addplot [
            color=RMSEColour2,
            dotted,
            line width=\plotLinewidth,
            no markers,
            forget plot,
        ]
            table [
                x=x_val,
                y=RSMA_sec_SOA_PRECODER_ONLY_PRIVATE_RATES_COUNTED_WSR_y,
                col sep=comma,
            ]{plots/rawData/TC2RsecOverCleaned.csv};
        
        \addplot [
            only marks,
            mark=triangle*,
            color=RMSEColour2,
            mark options={solid},
            mark size=\plotMarkersize,
            line width=\plotLinewidth,
            forget plot
        ]
            table [
                x=x_val,
                y=RSMA_sec_SOA_PRECODER_ONLY_PRIVATE_RATES_COUNTED_WSR_y,
                col sep=comma,
            ]{plots/markers/TC2Markers.csv};

        \addplot [
            color=RMSEColour8,
            dashed,
            line width=\plotLinewidth,
            no markers,
            forget plot,
        ]
            table [
                x=x_val,
                y=RSMAhyb_WSR_y,
                col sep=comma,
            ]{plots/rawData/TC2RsecOverCleaned.csv};
        
        \addplot [
            only marks,
            mark=x,
            color=RMSEColour8,
            mark options={solid},
            mark size=\plotMarkersize,
            line width=\plotLinewidth,
            forget plot,
        ]
            table [
                x=x_val,
                y=RSMAhyb_WSR_y,
                col sep=comma,
            ]{plots/markers/TC2Markers.csv};
        
        \coordinate (innerplot_upper_left) at (rel axis cs:0,1);
        \coordinate (innerplot_outer_corner) at (rel axis cs:1,1);
  
    \end{axis}

    \draw [
        thick,
        dotted,
        color=\zoomlinecolor
    ] (zoombox_lower_mid_left) -- (innerplot_upper_left);
    
    \draw [
        thick,
        dotted,
        color=\zoomlinecolor
    ] (zoombox_lower_mid_right) -- (innerplot_outer_corner);  
\end{tikzpicture}

%% file: plots/TC7_indiv_under_res.tex
\begin{tikzpicture}
    \def\xmin{0}
    \def\xmax{4.35}
    \def\ymin{-2}
    \def\ymax{4.6}
    \def\markrepeatrate{10}

	\begin{axis}[
		width=141, 
		height=120,
		xmin = \xmin, xmax = \xmax,
		ymin = \ymin, ymax = \ymax,
		xlabel={$R_{\text{s,1}}^{\mathrm{req}}$},
		ylabel={Partial Rates [bps/Hz]},
		grid=both,
		grid style={line width=.1pt, draw=gray!10},
		major grid style={line width=.2pt,draw=gray!50},
        legend style={
          font=\fontsize{6.5}{5}\selectfont,
          at={(0,0)},
          anchor=south west, 
          legend cell align=left,
          legend columns=2,
          cells={anchor=west},
          inner xsep=1.5pt,
          inner ysep=1.5pt,
          outer xsep=0pt,
          row sep=-2pt,
          column sep=1pt,
          mark size=1.4pt,
          mark options={solid},
        },
        legend image post style={xscale=0.77}, 
		ymode=linear,
        xtick={0,1,2,3,4},
        ytick={0,1,2,3,4},
        yticklabels={0,,2,,4},
		xlabel style={font=\fontsize{\fontsizevar}{\fontsizevar}\selectfont},
		ylabel style={font=\fontsize{\fontsizevarsmaller}{\fontsizevarsmaller}\selectfont},
		tick label style={font=\fontsize{\fontsizevar}{\fontsizevar}\selectfont},
		xlabel style={at={(axis description cs:0.5,-0.15)},rotate=0,anchor=south},
		ylabel style={at={(axis description cs:0.28,0.5)},rotate=0,anchor=south}
		]
		
		\addplot [mark=triangle*, 
        dotted, 
        color = RMSEColour1, 
        line width=0.8pt, 
        mark repeat={\markrepeatrate},
        mark options={solid}
        ]  table [
		x=x_val,
		y=RSMA_sec_SOA_PRECODER_ONLY_PRIVATE_RATES_COUNTED_com_y,
        col sep=comma,
		]{plots/rawData/TC5IndivRatesUnderCleaned.csv};

        \addplot [ mark=x, 
        dashed, 
        color = RMSEColour8, 
        line width=0.8pt, 
        mark repeat={\markrepeatrate},
        mark options={solid}
        ]  table [
		x=x_val,
		y=RSMAhyb_com_y,
		col sep=comma,
		]{plots/rawData/TC5IndivRatesUnderCleaned.csv};

        \addplot [
        mark=triangle*,
        dotted,
        color=RMSEColour3,
        line width=0.8pt,
        mark repeat={\markrepeatrate},
        mark options={solid}] table [
		x=x_val,
		y=RSMA_sec_SOA_PRECODER_ONLY_PRIVATE_RATES_COUNTED_priv1_y,
		col sep=comma,
		]{plots/rawData/TC5IndivRatesUnderCleaned.csv};

        \addplot [mark=x, 
        dashed, 
        color = RMSEColour6, 
        line width=0.8pt, 
        mark repeat={\markrepeatrate},
        mark options={solid}] table [
		x=x_val,
        y expr=\thisrow{RSMAhyb_priv1_y}-0.2,
		col sep=comma,
		]{plots/rawData/TC5IndivRatesUnderCleaned.csv};

        \addplot [mark=triangle*, 
        dotted, 
        color = RMSEColour2,
        line width=0.8pt,
        mark repeat={\markrepeatrate},
        mark options={solid}] table [
		x=x_val,
		y=RSMA_sec_SOA_PRECODER_ONLY_PRIVATE_RATES_COUNTED_priv2_y,
		col sep=comma,
		]{plots/rawData/TC5IndivRatesUnderCleaned.csv};

        \addplot [mark=x, 
        dashed, 
        color = RMSEColour5, 
        line width=0.8pt, 
        mark repeat={\markrepeatrate}, 
        mark options={solid},
        ] table [
		x=x_val,
        y expr=\thisrow{RSMAhyb_priv2_y}-0.1,
		col sep=comma,
		]{plots/rawData/TC5IndivRatesUnderCleaned.csv};	

		\legend{%
			Rpar(Cm),%
            Rhyb(Cm),%
			Rpar(Pr1),%
            Rhyb(Pr1),%
			Rpar(Pr2),%
            Rhyb(Pr2),%
		}
	\end{axis}
\end{tikzpicture}

%% file: plots/TC8_indiv_over_res.tex
\begin{tikzpicture}

    \def\zoomboxXmin{2.6}
    \def\zoomboxXmax{3.7}
    \def\zoomboxYmin{6.5}
    \def\zoomboxYmax{7.9}

    \pgfmathsetmacro{\zoomboxXmid}{0.5*(\zoomboxXmin+\zoomboxXmax)}
    \pgfmathsetmacro{\offsetmultiply}{0.5*(\zoomboxXmax-\zoomboxXmin)}
    \pgfmathsetmacro{\zoomboxlowermidleftx}{\zoomboxXmid - \offsetmultiply}
    \pgfmathsetmacro{\zoomboxlowermidrightx}{\zoomboxXmid + \offsetmultiply}
    \newcommand{\zoomboxLineWidth}{semithick}
    
    \def\zoomboxBorderColor{black}
    \def\zoomboxBgColor{none}
    \def\zoomboxOpacity{1}
    \def\zoomlinecolor{black}

    \def\arrowyStart{7.183212698451356}
    \def\arrowyEnd{7.083212698451356}

    \def\markersecondlast{2.8000000000000002665}

    \def\markrepeatrate{8}
    \def\maskingpoint{0.5}
    \def\xmin{0}
    \def\xmin{0}
    \def\xmax{4.35}
    \def\ymin{-2.9}
    \def\ymax{10.7}
    \def\markrepeatrate{10}

    \def\TCEightLegendLinewidth{0.77pt}
    \def\TCEightLegendMarkSize{2.2pt}
    
	\begin{axis}[
		width=141, 
		height=120,
		xmin = \xmin, xmax = \xmax,
		ymin = \ymin, ymax = \ymax,
		xlabel={$R_{\text{s},1}^{\mathrm{req}}$},
		grid=both,
		grid style={line width=.1pt, draw=gray!10},
		major grid style={line width=.2pt,draw=gray!50},
        legend style={
          font=\fontsize{6.1}{5}\selectfont,
          at={(0,0)},
          anchor=south west, 
          legend cell align=left,
          legend columns=2,
          cells={anchor=west},
          inner xsep=2.5pt,
          inner ysep=1pt,
          outer xsep=0pt,
          row sep=-2pt,
          mark size=\TCEightLegendMarkSize,
          mark options={solid},
        },
        legend image post style={xscale=0.420}, 
		ymode=linear,
        ytick={0,2,4,6,8,10},
        yticklabels={0,,4,,8,,},
        xtick={0,1,2,3,4},
		xlabel style={font=\fontsize{\fontsizevar}{\fontsizevar}\selectfont},
		tick label style={font=\fontsize{\fontsizevar}{\fontsizevar}\selectfont},
		xlabel style={at={(axis description cs:0.5,-0.15)},rotate=0,anchor=south},
		]

        \addplot [
            color = RMSEColour4, 
            line width=0.8pt, 
            no markers, 
            forget plot
        ]  table [
    		x=x_val,
    		y=SDMA_priv1_y,
    		col sep=comma,
		]{plots/rawData/TC6IndivRatesOverCleaned.csv};

        \addplot [
            only marks,
            mark=diamond,
            color=RMSEColour4,
            mark size=\plotMarkersize,
            line width=\plotLinewidth,
            forget plot,
        ]table [
            x=x_val,
            y=SDMA_priv1_y,
            col sep=comma,
        ]{plots/markers/TC6Markers.csv};

        \addplot [
            color = RMSEColour7, 
            line width=0.8pt, 
            no markers, 
            forget plot
        ]  table [
    		x=x_val,
    		y=SDMA_priv2_y,
    		col sep=comma,
		]{plots/rawData/TC6IndivRatesOverCleaned.csv};

        \addplot [
            only marks,
            mark=diamond,
            color=RMSEColour7,
            mark size=\plotMarkersize,
            line width=\plotLinewidth,
            forget plot,
        ]table [
            x=x_val,
            y=SDMA_priv2_y,
            col sep=comma,
        ]{plots/markers/TC6Markers.csv};

        \addplot [
            dotted,
            color = RMSEColour1, 
            line width=0.8pt, 
            no markers, 
            forget plot
        ]  table [
    		x=x_val,
    		y=RSMA_sec_SOA_PRECODER_ONLY_PRIVATE_RATES_COUNTED_com_y,
    		col sep=comma,
		]{plots/rawData/TC6IndivRatesOverCleaned.csv};

        \addplot [
            only marks,
            mark=triangle*,
            color=RMSEColour1,
            mark size=\plotMarkersize,
            line width=\plotLinewidth,
            forget plot,
        ]table [
            x=x_val,
            y=RSMA_sec_SOA_PRECODER_ONLY_PRIVATE_RATES_COUNTED_com_y,
            col sep=comma,
        ]{plots/markers/TC6Markers.csv};
        
        \addplot [
            dotted,
            color = RMSEColour3, 
            line width=0.8pt, 
            no markers, 
            forget plot
        ]  table [
    		x=x_val,
    		y=RSMA_sec_SOA_PRECODER_ONLY_PRIVATE_RATES_COUNTED_priv1_y,
    		col sep=comma,
		]{plots/rawData/TC6IndivRatesOverCleaned.csv};

        \addplot [
            only marks,
            mark=triangle*,
            color=RMSEColour3,
            mark size=\plotMarkersize,
            line width=\plotLinewidth,
            forget plot,
        ]table [
            x=x_val,
            y=RSMA_sec_SOA_PRECODER_ONLY_PRIVATE_RATES_COUNTED_priv1_y,
            col sep=comma,
        ]{plots/markers/TC6Markers.csv};

        \addplot [
            dotted,
            color = RMSEColour2, 
            line width=0.8pt, 
            no markers, 
            forget plot
        ]  table [
    		x=x_val,
    		y=RSMA_sec_SOA_PRECODER_ONLY_PRIVATE_RATES_COUNTED_priv2_y,
    		col sep=comma,
		]{plots/rawData/TC6IndivRatesOverCleaned.csv};

        \addplot [
            only marks,
            mark=triangle*,
            color=RMSEColour2,
            mark size=\plotMarkersize,
            line width=\plotLinewidth,
            forget plot,
        ]table [
            x=x_val,
            y=RSMA_sec_SOA_PRECODER_ONLY_PRIVATE_RATES_COUNTED_priv2_y,
            col sep=comma,
        ]{plots/markers/TC6Markers.csv};

        \addplot [
            dashed,
            color = RMSEColour8, 
            line width=0.8pt, 
            no markers, 
            forget plot
        ]  table [
    		x=x_val,
    		y=RSMAhyb_com_y,
    		col sep=comma,
		]{plots/rawData/TC6IndivRatesOverCleaned.csv};

        \addplot [
            only marks,
            mark=x,
            color=RMSEColour8,
            mark size=\plotMarkersize,
            line width=\plotLinewidth,
            forget plot,
        ]table [
            x=x_val,
            y=RSMAhyb_com_y,
            col sep=comma,
        ]{plots/markers/TC6Markers.csv};
        
        \addplot [
            dashed,
            color = RMSEColour6, 
            line width=0.8pt, 
            no markers, 
            forget plot
        ]  table [
		      x=x_val,
            y expr=\thisrow{RSMAhyb_priv1_y}-0.2,
    		col sep=comma,
		]{plots/rawData/TC6IndivRatesOverCleaned.csv};

        \addplot [
            only marks,
            mark=x,
            color=RMSEColour6,
            mark size=\plotMarkersize,
            line width=\plotLinewidth,
            forget plot,
        ]table [
    		x=x_val,
            y expr=\thisrow{RSMAhyb_priv1_y}-0.2,
    		col sep=comma,
        ]{plots/markers/TC6Markers.csv};

        \addplot [
            dashed,
            color = RMSEColour5, 
            line width=0.8pt, 
            no markers, 
            forget plot
        ]  table [
		      x=x_val,
            y expr=\thisrow{RSMAhyb_priv2_y}-0.1,
    		col sep=comma,
		]{plots/rawData/TC6IndivRatesOverCleaned.csv};

        \addplot [
            only marks,
            mark=x,
            color=RMSEColour5,
            mark size=\plotMarkersize,
            line width=\plotLinewidth,
            forget plot,
        ]table [
    		x=x_val,
            y expr=\thisrow{RSMAhyb_priv2_y}-0.1,
    		col sep=comma,
        ]{plots/markers/TC6Markers.csv};

        \addlegendimage{color=RMSEColour4,mark=diamond,mark options={solid, xscale=1.6, yscale=1}, line width=\TCEightLegendLinewidth,mark size=\TCEightLegendMarkSize}
        \addlegendentry{SSDMA(Pr1)}
    
        \addlegendimage{color=RMSEColour7,mark=diamond,mark options={solid, xscale=1.6, yscale=1},line width=\TCEightLegendLinewidth,mark size=\TCEightLegendMarkSize}
        \addlegendentry{SSDMA(Pr2)}

        \addplot [
            draw=\zoomboxBorderColor,
            fill=\zoomboxBgColor,
            opacity=\zoomboxOpacity,
            \zoomboxLineWidth
        ] 
        coordinates {
            (\zoomboxXmin,\zoomboxYmin)
            (\zoomboxXmax,\zoomboxYmin)
            (\zoomboxXmax,\zoomboxYmax)
            (\zoomboxXmin,\zoomboxYmax)
            (\zoomboxXmin,\zoomboxYmin)
        };
        \coordinate (zoombox_lower_mid_left) at (axis cs:\zoomboxlowermidleftx,\zoomboxYmin);
        \coordinate (zoombox_lower_mid_right) at (axis cs:\zoomboxlowermidrightx,\zoomboxYmin);
		
	\end{axis}

    \begin{axis}[
        name=subplot,
        at={(mainplot.south east)},
        axis lines=box,
        width=89, 
		height=70,
        xshift=-3.5pt,
        yshift=47.2pt,
		grid style={line width=.1pt, draw=gray!10},
		major grid style={line width=.2pt,draw=gray!50},
        anchor=north east,
        grid=both,
        xtick={3},
        xticklabels={},
        ytick={7.1, 7.2},
        xmin=\zoomboxXmin, xmax=\zoomboxXmax,
        ymin=7.04, ymax=7.2315,
        xlabel style={font=\fontsize{\tiksizeVarInnerPlot}{\tiksizeVarInnerPlot}\selectfont},
		ylabel style={font=\fontsize{\tiksizeVarInnerPlot}{\tiksizeVarInnerPlot}\selectfont},
		tick label style={font=\fontsize{\tiksizeVarInnerPlot}{\tiksizeVarInnerPlot}\selectfont},
        axis background/.style={fill=white},
        axis line style={\zoomboxLineWidth, draw=\zoomboxBorderColor}
       ]

        \addplot [
            color = RMSEColour7, 
            line width=0.8pt, 
            no markers, 
            forget plot
        ]  table [
    		x=x_val,
    		y=SDMA_priv2_y,
    		col sep=comma,
		]{plots/rawData/TC6IndivRatesOverCleaned.csv};

        \addplot [
            only marks,
            mark=diamond*,
            color=RMSEColour7,
            mark size=\plotMarkersize,
            line width=\plotLinewidth, 
            forget plot,
        ]table [
            x=x_val,
            y=SDMA_priv2_y,
            col sep=comma,
        ]{plots/markers/TC6Markers.csv};

        \addplot [
            dotted,
            color = RMSEColour2, 
            line width=0.8pt, 
            no markers, 
            forget plot
        ]  table [
    		x=x_val,
    		y=RSMA_sec_SOA_PRECODER_ONLY_PRIVATE_RATES_COUNTED_priv2_y,
    		col sep=comma,
		]{plots/rawData/TC6IndivRatesOverCleaned.csv};

        \addplot [
            only marks,
            mark=triangle*,
            color=RMSEColour2,
            mark size=\plotMarkersize,
            line width=\plotLinewidth,
            forget plot,
        ]table [
            x=x_val,
            y=RSMA_sec_SOA_PRECODER_ONLY_PRIVATE_RATES_COUNTED_priv2_y,
            col sep=comma,
        ]{plots/markers/TC6Markers.csv};

        \addplot [
            dashed,
            color = RMSEColour5, 
            line width=0.8pt, 
            no markers, 
            forget plot
        ]  table [
		      x=x_val,
            y expr=\thisrow{RSMAhyb_priv2_y}-0.1,
    		col sep=comma,
		]{plots/rawData/TC6IndivRatesOverCleaned.csv};

        \addplot [
            only marks,
            mark=x,
            color=RMSEColour5,
            mark size=\plotMarkersize,
            line width=\plotLinewidth,
            forget plot,
        ]table [
    		x=x_val,
            y expr=\thisrow{RSMAhyb_priv2_y}-0.1,
    		col sep=comma,
        ]{plots/markers/TC6Markers.csv};

        \coordinate (innerplot_upper_left) at (rel axis cs:0,1);
        \coordinate (innerplot_outer_corner) at (rel axis cs:1,1);

        \draw[{Stealth[scale=0.6]}-{Stealth[scale=0.6]}, dotted, color=black!80, thick, shorten <=2.3pt]
        (axis cs:\markersecondlast,\arrowyStart)
        -- (axis cs:\markersecondlast,\arrowyEnd)
        node[midway, right, yshift=-0.65pt, xshift=1.6pt, scale=0.6, transform shape] {$R^{\mathrm{req}}_{\text{i},2} \negVeryTiny =  \negVeryTiny \negVeryTiny 0.1$};
  
    \end{axis}
    
        \draw [
            thick,
            dotted,
            color=\zoomlinecolor
        ] (zoombox_lower_mid_left) -- (innerplot_upper_left);
        
        \draw [
            thick,
            dotted,
            color=\zoomlinecolor
        ] (zoombox_lower_mid_right) -- (innerplot_outer_corner);  
\end{tikzpicture}

%% file: plots/TC9_Rsec_rnd_under.tex
\begin{tikzpicture}

    \def\TCNineLegendLinewidth{0.8pt}
    \def\TCNineLegendMarkSize{1.9pt}
    
	\begin{axis}[
		width=143, 
		height=120,
		xmin = 0, xmax = 4.6,
		ymin = -0.1, ymax = 13.1,
		xlabel={$R_{\text{s},1}^{\mathrm{req}}$},
		ylabel={$\overline{\text{SE}}$ [bps/Hz]},
		grid=both,
		grid style={line width=.1pt, draw=gray!10},
		major grid style={line width=.2pt,draw=gray!50},
        legend style={
          font=\fontsize{7.3}{7.3}\selectfont,
          at={(0,0)},
          anchor=south west, 
          legend columns=1,
          cells={anchor=west},
          inner xsep=2.5pt,
          inner ysep=1pt,
          row sep=-2pt,
          scale=0.5,
          mark size=1.5pt,
          line width=0.3pt
        },
		ymode=linear,
        ytick={0,3,6,9,12},
        xtick={0,1,2,3,4},
		xlabel style={font=\fontsize{\fontsizevar}{\fontsizevar}\selectfont},
		ylabel style={font=\fontsize{\fontsizevarsYLabel}{\fontsizevarsYLabel}\selectfont},
		tick label style={font=\fontsize{\fontsizevar}{\fontsizevar}\selectfont},
		xlabel style={at={(axis description cs:0.5,-0.15)},rotate=0,anchor=south},
		ylabel style={at={(axis description cs:0.27,0.5)},rotate=0,anchor=south}
		]
        \def\markrepeatrate{6}

        \addplot [
        mark=diamond,
        color=RMSEColour3,
        line width=0.8pt, mark repeat={\markrepeatrate}, 
        mark options={solid},
        forget plot
        ] table [
		x=x_val,
		y=SDMA_WSR_y,
		col sep=comma,
		]{plots/rawData/TC9RsecRndUnderCleaned.csv};   

        \addplot [mark=triangle*, 
        dotted,
        color = RMSEColour2, 
        line width=0.8pt, 
        mark repeat={\markrepeatrate},mark options={solid},
        forget plot
        ] table [
		x=x_val,
		y=RSMA_sec_SOA_PRECODER_ONLY_PRIVATE_RATES_COUNTED_WSR_y,
		col sep=comma,
		]{plots/rawData/TC9RsecRndUnderCleaned.csv};

        \addplot [ mark=x, 
        dashed,
        color = RMSEColour8, 
        line width=0.8pt, 
        mark repeat={\markrepeatrate},
        mark options={solid},
        forget plot
        ] table [
		x=x_val,
		y=RSMAhyb_WSR_y,
		col sep=comma,
		]{plots/rawData/TC9RsecRndUnderCleaned.csv};
		
        \addlegendimage{color=RMSEColour3, mark=diamond,mark options={solid, xscale=1.1, yscale=1}, line width=\TCNineLegendLinewidth, mark size=\TCNineLegendMarkSize}
        \addlegendentry{SSDMA}

        \addlegendimage{color=RMSEColour2, dotted, mark=triangle*, mark options={solid, xscale=1.1, yscale=1}, line width=\TCNineLegendLinewidth, mark size=\TCNineLegendMarkSize}
        \addlegendentry{Rpar}

        \addlegendimage{color=RMSEColour8, dashed, mark=x, mark options={solid, xscale=1.1, yscale=1}, line width=\TCNineLegendLinewidth, mark size=\TCNineLegendMarkSize}
        \addlegendentry{Rhyb}
	\end{axis}
\end{tikzpicture}

%% file: plots/TC10_Rsec_rnd_over.tex
\begin{tikzpicture}
    \def\zoomboxXmin{2.8}
    \def\zoomboxXmax{3.8}
    \def\zoomboxYmin{16.5}
    \def\zoomboxYmax{19.5}

    \pgfmathsetmacro{\zoomboxXmid}{0.5*(\zoomboxXmin+\zoomboxXmax)}
    \pgfmathsetmacro{\offsetmultiply}{0.5*(\zoomboxXmax-\zoomboxXmin)}
    \pgfmathsetmacro{\zoomboxlowermidleftx}{\zoomboxXmid - \offsetmultiply}
    \pgfmathsetmacro{\zoomboxlowermidrightx}{\zoomboxXmid + \offsetmultiply}
    \newcommand{\zoomboxLineWidth}{semithick}

    \def\zoomboxBorderColor{black}
    \def\zoomboxBgColor{none}
    \def\zoomboxOpacity{1}
    \def\zoomlinecolor{black}

    \def\legendLinewidth{0.8pt}

	\begin{axis}[
		width=143, 
		height=120,
		xmin = 0, xmax = 4.6,
		ymin = -0.1, ymax = 22,
		xlabel={$R_{\text{s},1}^{\mathrm{req}}$},
		grid=both,
		grid style={line width=.1pt, draw=gray!10},
		major grid style={line width=.2pt,draw=gray!50},
        legend style={
          font=\fontsize{7.1}{7.1}\selectfont,
          at={(0,0)},
          anchor=south west,
          legend columns=2,
          cells={anchor=west},
          inner xsep=2.5pt,
          inner ysep=1pt,
          row sep=-2pt,
          scale=0.5,
          mark size=1.2pt,
          line width=0.3pt
        },
		ymode=linear,
        ytick={0, 5, 10, 15, 20},
        xtick={0,1,2,3,4},
		xlabel style={font=\fontsize{\fontsizevar}{\fontsizevar}\selectfont},
		ylabel style={font=\fontsize{\fontsizevar}{\fontsizevar}\selectfont},
		tick label style={font=\fontsize{\fontsizevar}{\fontsizevar}\selectfont},
		xlabel style={at={(axis description cs:0.5,-0.15)},rotate=0,anchor=south},
		ylabel style={at={(axis description cs:0.3,0.5)},rotate=0,anchor=south}
		]

        \addplot [
            color=RMSEColour3,
            line width=\plotLinewidth,
            no markers,
            forget plot
        ]
            table [
                x=x_val,
                y=SDMA_WSR_y,
                col sep=comma,
            ]{plots/rawData/TC10RsecRndOverCleaned.csv};
        
        \addplot [
            only marks,
            mark=diamond,
            color=RMSEColour3,
            mark size=\plotMarkersize,
            line width=\plotLinewidth,
            forget plot
        ]
            table [
                x=x_val,
                y=SDMA_WSR_y,
                col sep=comma,
            ]{plots/markers/TC10Markers.csv};

        \addplot [
            color=RMSEColour2,
            dotted,
            line width=\plotLinewidth,
            no markers,
            forget plot
        ]
            table [
                x=x_val,
                y=RSMA_sec_SOA_PRECODER_ONLY_PRIVATE_RATES_COUNTED_WSR_y,
                col sep=comma,
            ]{plots/rawData/TC10RsecRndOverCleaned.csv};
        
        \addplot [
            only marks,
            mark=triangle*,
            color=RMSEColour2,
            mark options={solid},
            mark size=\plotMarkersize,
            line width=\plotLinewidth,
            forget plot
        ]
            table [
                x=x_val,
                y=RSMA_sec_SOA_PRECODER_ONLY_PRIVATE_RATES_COUNTED_WSR_y,
                col sep=comma,
            ]{plots/markers/TC10Markers.csv};

        \addplot [
            color=RMSEColour8,
            dashed,
            line width=\plotLinewidth,
            no markers,
            forget plot
        ]
            table [
                x=x_val,
                y=RSMAhyb_WSR_y,
                col sep=comma,
            ]{plots/rawData/TC10RsecRndOverCleaned.csv};
        
        \addplot [
            only marks,
            mark=x,
            color=RMSEColour8,
            mark options={solid},
            mark size=\plotMarkersize,
            line width=\plotLinewidth,
            forget plot
        ]
            table [
                x=x_val,
                y=RSMAhyb_WSR_y,
                col sep=comma,
            ]{plots/markers/TC10Markers.csv};

        \addplot [
            draw=\zoomboxBorderColor,
            fill=\zoomboxBgColor,
            opacity=\zoomboxOpacity,
            \zoomboxLineWidth
        ] 
        coordinates {
            (\zoomboxXmin,\zoomboxYmin)
            (\zoomboxXmax,\zoomboxYmin)
            (\zoomboxXmax,\zoomboxYmax)
            (\zoomboxXmin,\zoomboxYmax)
            (\zoomboxXmin,\zoomboxYmin)
        };
    \coordinate (zoombox_lower_mid_left) at (axis cs:\zoomboxlowermidleftx,\zoomboxYmin);
    \coordinate (zoombox_lower_mid_right) at (axis cs:\zoomboxlowermidrightx,\zoomboxYmin);
	\end{axis}

    \begin{axis}[
        name=subplot,
        at={(mainplot.south east)},
        axis lines=box,
        width=80, 
		height=82,
        xshift=-35pt,
        yshift=46.8pt,
		grid style={line width=.1pt, draw=gray!10},
		major grid style={line width=.2pt,draw=gray!50},
        anchor=north east,
        grid=both,
        xtick={3},
        xticklabels={3},
        extra x ticks={3.5},
        extra x tick labels={3.5},
        extra x tick style={
          tick style={draw=none},
          grid=none
        },
        ytick={17.3, 17.7, 18.1},
        xmin=2.8, xmax=3.8,
        ymin=17.2, ymax=18.3,
        xlabel style={font=\fontsize{\tiksizeVarInnerPlot}{\tiksizeVarInnerPlot}\selectfont},
		ylabel style={font=\fontsize{\tiksizeVarInnerPlot}{\tiksizeVarInnerPlot}\selectfont},
		tick label style={font=\fontsize{\tiksizeVarInnerPlot}{\tiksizeVarInnerPlot}\selectfont},
        axis background/.style={fill=white},
        axis line style={\zoomboxLineWidth, draw=\zoomboxBorderColor}
       ]

        \addplot [
            color=RMSEColour3,
            line width=\plotLinewidth,
            no markers,
            forget plot,
        ]
            table [
                x=x_val,
                y=SDMA_WSR_y,
                col sep=comma,
            ]{plots/rawData/TC10RsecRndOverCleaned.csv};
        
        \addplot [
            only marks,
            mark=diamond*,
            color=RMSEColour3,
            mark size=\plotMarkersize,
            line width=\plotLinewidth,
            forget plot,
        ]
            table [
                x=x_val,
                y=SDMA_WSR_y,
                col sep=comma,
            ]{plots/markers/TC10Markers.csv};

        \addplot [
            color=RMSEColour2,
            dotted,
            line width=\plotLinewidth,
            no markers,
            forget plot,
        ]
            table [
                x=x_val,
                y=RSMA_sec_SOA_PRECODER_ONLY_PRIVATE_RATES_COUNTED_WSR_y,
                col sep=comma,
            ]{plots/rawData/TC10RsecRndOverCleaned.csv};
        
        \addplot [
            only marks,
            mark=triangle*,
            color=RMSEColour2,
            mark options={solid},
            mark size=\plotMarkersize,
            line width=\plotLinewidth,
            forget plot
        ]
            table [
                x=x_val,
                y=RSMA_sec_SOA_PRECODER_ONLY_PRIVATE_RATES_COUNTED_WSR_y,
                col sep=comma,
            ]{plots/markers/TC10Markers.csv};

        \addplot [
            color=RMSEColour8,
            dashed,
            line width=\plotLinewidth,
            no markers,
            forget plot,
        ]
            table [
                x=x_val,
                y=RSMAhyb_WSR_y,
                col sep=comma,
            ]{plots/rawData/TC10RsecRndOverCleaned.csv};
        
        \addplot [
            only marks,
            mark=x,
            color=RMSEColour8,
            mark options={solid},
            mark size=\plotMarkersize,
            line width=\plotLinewidth,
            forget plot,
        ]
            table [
                x=x_val,
                y=RSMAhyb_WSR_y,
                col sep=comma,
            ]{plots/markers/TC10Markers.csv};      
        \coordinate (innerplot_upper_left) at (rel axis cs:0,1);
        \coordinate (innerplot_outer_corner) at (rel axis cs:1,1);
  
    \end{axis}
    
    \draw [
        thick,
        dotted,
        color=\zoomlinecolor
    ] (zoombox_lower_mid_left) -- (innerplot_upper_left);
    
    \draw [
        thick,
        dotted,
        color=\zoomlinecolor
    ] (zoombox_lower_mid_right) -- (innerplot_outer_corner);  
\end{tikzpicture}

%% file: plots/TC11_Rsec_rnd_over_infeas.tex
\begin{tikzpicture}
	\begin{axis}[
		width=141,  
		height=120,
		xmin = 0, xmax = 4.6,
		ymin = 0, ymax = 110,
		xlabel={$R_{\text{s},1}^{\mathrm{req}}$},
		ylabel={Unsolvable [\%]},
		grid=both,
		grid style={line width=.1pt, draw=gray!10},
		major grid style={line width=.2pt,draw=gray!50},
        legend style={
          font=\fontsize{7.1}{7.1}\selectfont,
          at={(0,0)},
          anchor=south west,
          legend columns=2,
          cells={anchor=west},
          inner xsep=2.5pt,
          inner ysep=1pt,
          row sep=-2pt,
          scale=0.5,
          mark size=1.2pt,
          line width=0.3pt
        },
		ymode=linear,
        ytick={0, 25, 50, 75, 100},
        xtick={0,1,2,3,4},
		xlabel style={font=\fontsize{\fontsizevar}{\fontsizevar}\selectfont},
		ylabel style={font=\fontsize{\fontsizevarsYLabelTC11}{\fontsizevarsYLabelTC11}\selectfont},
		tick label style={font=\fontsize{\fontsizevar}{\fontsizevar}\selectfont},
		xlabel style={at={(axis description cs:0.5,-0.15)},rotate=0,anchor=south},
		ylabel style={at={(axis description cs:0.22,0.5)},rotate=0,anchor=south}
		]
        \def\markrepeatrate{6}

        \addplot [
        mark=diamond,
        color=RMSEColour3,
        line width=0.8pt, mark repeat={\markrepeatrate}] table [
		x=x_val,
		col sep=comma,
        y expr=(\thisrow{SDMA_infeas_y}/2500)*100,
		]{plots/rawData/TC9RsecRndUnderInfeasCleaned.csv};   

        \addplot [mark=triangle*,
        dotted,
        color = RMSEColour2, 
        line width=0.8pt, 
        mark repeat={\markrepeatrate}] table [
		x=x_val,
        y expr=(\thisrow{RSMA_sec_SOA_PRECODER_ONLY_PRIVATE_RATES_COUNTED_infeas_y}/2500)*100,
		col sep=comma,
		]{plots/rawData/TC9RsecRndUnderInfeasCleaned.csv};

        \addplot [ mark=x, 
        dashed,
        color = RMSEColour8, 
        line width=0.8pt, 
        mark repeat={\markrepeatrate},]  table [
		x=x_val,
        y expr=(\thisrow{RSMAhyb_infeas_y}/2500)*100,
		col sep=comma,
		]{plots/rawData/TC9RsecRndUnderInfeasCleaned.csv};

	\end{axis}
\end{tikzpicture}

%% file: plots/TC12_Rsec_rnd_over_infeas.tex
\begin{tikzpicture}
    \def\zoomboxXmin{2.55}
    \def\zoomboxXmax{3.75}
    \def\zoomboxYmin{0}
    \def\zoomboxYmax{5}

    \pgfmathsetmacro{\zoomboxXmid}{0.5*(\zoomboxXmin+\zoomboxXmax)}
    \pgfmathsetmacro{\offsetmultiply}{0.5*(\zoomboxXmax-\zoomboxXmin)}
    \pgfmathsetmacro{\zoomboxlowermidleftx}{\zoomboxXmid - \offsetmultiply}
    \pgfmathsetmacro{\zoomboxlowermidrightx}{\zoomboxXmid + \offsetmultiply}
    \newcommand{\zoomboxLineWidth}{semithick}
    
    \def\zoomboxBorderColor{black}
    \def\zoomboxBgColor{none}
    \def\zoomboxOpacity{1}
    \def\zoomlinecolor{black}

	\begin{axis}[
		width=141, 
		height=120,
		xmin = 0, xmax = 4.6,
		ymin = 0, ymax = 110,
		xlabel={$R_{\text{s},1}^{\mathrm{req}}$},
		grid=both,
		grid style={line width=.1pt, draw=gray!10},
		major grid style={line width=.2pt,draw=gray!50},
        legend style={
          font=\fontsize{7.1}{7.1}\selectfont,
          at={(0,0)},
          anchor=south west,
          legend columns=2,
          cells={anchor=west},
          inner xsep=2.5pt,
          inner ysep=1pt,
          row sep=-2pt,
          scale=0.5,
          mark size=1.2pt,
          line width=0.3pt
        },
		ymode=linear,
        ytick={0, 25, 50, 75, 100},
        xtick={0,1,2,3,4},
		xlabel style={font=\fontsize{\fontsizevar}{\fontsizevar}\selectfont},
		ylabel style={font=\fontsize{\fontsizevar}{\fontsizevar}\selectfont},
		tick label style={font=\fontsize{\fontsizevar}{\fontsizevar}\selectfont},
		xlabel style={at={(axis description cs:0.5,-0.15)},rotate=0,anchor=south},
		ylabel style={at={(axis description cs:0.3,0.5)},rotate=0,anchor=south}
		]
        \def\markrepeatrate{6}

        \addplot [
            mark=diamond, 
            color = RMSEColour3, 
            line width=0.8pt, 
            no markers, 
            forget plot
        ]  table [
    		x=x_val,
    		y expr=(\thisrow{SDMA_infeas_y}/2500)*100,
    		col sep=comma,
		]{plots/rawData/TC10RsecRndOverInfeasCleaned.csv};

        \addplot [
            only marks,
            mark=diamond,
            color=RMSEColour3,
            mark size=\plotMarkersize,
            line width=\plotLinewidth,
            forget plot,
        ]table [
            x=x_val,
            y expr=(\thisrow{SDMA_infeas_y}/2500)*100,
            col sep=comma,
        ]{plots/markers/TC10InfeasMarkers.csv};   
        
        \addplot [
            dotted,
            color = RMSEColour2, 
            line width=0.8pt, 
            no markers, 
            forget plot
        ]  table [
    		x=x_val,
    		y expr=(\thisrow{RSMA_sec_SOA_PRECODER_ONLY_PRIVATE_RATES_COUNTED_infeas_y}/2500)*100,
    		col sep=comma,
		]{plots/rawData/TC10RsecRndOverInfeasCleaned.csv};

        \addplot [
            only marks,
            mark=triangle*,
            color=RMSEColour2,
            mark size=\plotMarkersize,
            line width=\plotLinewidth,
            forget plot,
        ]table [
            x=x_val,
            y    expr=(\thisrow{RSMA_sec_SOA_PRECODER_ONLY_PRIVATE_RATES_COUNTED_infeas_y}/2500)*100,
            col sep=comma,
        ]{plots/markers/TC10InfeasMarkers.csv};

        \addplot [
            dashed,
            color = RMSEColour8, 
            line width=0.8pt, 
            no markers, 
            forget plot
        ]  table [
    		x=x_val,
    		y expr=(\thisrow{RSMAhyb_infeas_y}/2500)*100,
    		col sep=comma,
		]{plots/rawData/TC10RsecRndOverInfeasCleaned.csv};

        \addplot [
            dashed,
            only marks,
            mark=x,
            color=RMSEColour8,
            mark size=\plotMarkersize,
            line width=\plotLinewidth,
            forget plot,
        ]table [
            x=x_val,
            y expr=(\thisrow{RSMAhyb_infeas_y}/2500)*100,
            col sep=comma,
        ]{plots/markers/TC10InfeasMarkers.csv};      	

        \addplot [
            draw=\zoomboxBorderColor,
            fill=\zoomboxBgColor,
            opacity=\zoomboxOpacity,
            \zoomboxLineWidth
        ] 
        coordinates {
            (\zoomboxXmin,\zoomboxYmin)
            (\zoomboxXmax,\zoomboxYmin)
            (\zoomboxXmax,\zoomboxYmax)
            (\zoomboxXmin,\zoomboxYmax)
            (\zoomboxXmin,\zoomboxYmin)
        };

        \coordinate (zoombox_upper_mid_left)  at (axis cs:\zoomboxlowermidleftx,\zoomboxYmax);
        \coordinate (zoombox_upper_mid_right) at (axis cs:\zoomboxlowermidrightx,\zoomboxYmax);

	\end{axis}

    \begin{axis}[
        name=subplot,
        at={(mainplot.south east)},
        axis lines=box,
        width=80, 
		height=60,
        xshift=-48.5pt,
        yshift=50.1,
		grid style={line width=.1pt, draw=gray!10},
		major grid style={line width=.2pt,draw=gray!50},
        anchor=north east,
        grid=both,
        xtick={3},
        xticklabels={3},
        extra x ticks={3.5},
        extra x tick labels={3.5},
        extra x tick style={
          tick style={draw=none},
          grid=none
        },
        xtick pos=top,
        ytick={0},
        xmin=\zoomboxXmin, xmax=\zoomboxXmax+0.1, 
        ymin=-0.5, ymax=1,
        xlabel style={font=\fontsize{\tiksizeVarInnerPlot}{\tiksizeVarInnerPlot}\selectfont},
		ylabel style={font=\fontsize{\tiksizeVarInnerPlot}{\tiksizeVarInnerPlot}\selectfont},
		tick label style={font=\fontsize{\tiksizeVarInnerPlot}{\tiksizeVarInnerPlot}\selectfont},
        axis background/.style={fill=white},
        axis line style={\zoomboxLineWidth, draw=\zoomboxBorderColor}
       ]
       
        \addplot [
            mark=diamond, 
            color = RMSEColour3, 
            line width=0.8pt, 
            no markers, 
            forget plot
        ]  table [
    		x=x_val,
    		y expr=(\thisrow{SDMA_infeas_y}/2500)*100,
    		col sep=comma,
		]{plots/rawData/TC10RsecRndOverInfeasCleaned.csv};

        \addplot [
            only marks,
            mark=diamond,
            color=RMSEColour3,
            mark size=\plotMarkersize,
            line width=\plotLinewidth,
            forget plot,
        ]table [
            x=x_val,
            y expr=(\thisrow{SDMA_infeas_y}/2500)*100,
            col sep=comma,
        ]{plots/markers/TC10InfeasMarkers.csv};
        
        \addplot [
            dotted,
            color = RMSEColour2, 
            line width=0.8pt, 
            no markers, 
            forget plot
        ]  table [
    		x=x_val,
    		y expr=(\thisrow{RSMA_sec_SOA_PRECODER_ONLY_PRIVATE_RATES_COUNTED_infeas_y}/2500)*100,
    		col sep=comma,
		]{plots/rawData/TC10RsecRndOverInfeasCleaned.csv};

        \addplot [
            only marks,
            mark=triangle*,
            color=RMSEColour2,
            mark size=\plotMarkersize,
            line width=\plotLinewidth,
            forget plot,
        ]table [
            x=x_val,
            y    expr=(\thisrow{RSMA_sec_SOA_PRECODER_ONLY_PRIVATE_RATES_COUNTED_infeas_y}/2500)*100,
            col sep=comma,
        ]{plots/markers/TC10InfeasMarkers.csv};

        \addplot [
            dashed,
            color = RMSEColour8, 
            line width=0.8pt, 
            no markers, 
            forget plot
        ]  table [
    		x=x_val,
    		y expr=(\thisrow{RSMAhyb_infeas_y}/2500)*100,
    		col sep=comma,
		]{plots/rawData/TC10RsecRndOverInfeasCleaned.csv};

        \addplot [
            dashed,
            only marks,
            mark=x,
            color=RMSEColour8,
            mark size=\plotMarkersize,
            line width=\plotLinewidth,
            forget plot,
        ]table [
            x=x_val,
            y expr=(\thisrow{RSMAhyb_infeas_y}/2500)*100,
            col sep=comma,
        ]{plots/markers/TC10InfeasMarkers.csv};
        
        \coordinate (innerplot_lower_left) at (rel axis cs:0,0);
        \coordinate (innerplot_lower_right) at (rel axis cs:1,0);
    \end{axis}

    \draw [
        thick,
        dotted,
        color=\zoomlinecolor
    ] (zoombox_upper_mid_left) -- (innerplot_lower_left);
    
    \draw [
        thick,
        dotted,
        color=\zoomlinecolor
    ] (zoombox_upper_mid_right) -- (innerplot_lower_right);  
\end{tikzpicture}

%% file: chapters/06_Conclusion.tex
\section{Conclusion}
\label{sec:concl}
In this work, we presented a hybrid-\ac{PLS} \ac{RSMA} system that symmetrically encrypts user data and transmits the related key information within the \ac{PLS}-secured private message. To facilitate this system, a dedicated \ac{RSMA} precoder was formulated to ensure the required size and security of the private messages. The precoder design was cast as a non-convex \ac{SE} maximization problem and solved using an \ac{SCA}-based algorithm. Analytical results for the \ac{FSDT} scenario establish that hybrid-\ac{PLS} \ac{RSMA} significantly increases \ac{SE} in the \ac{NOMA} operational region, whereas in the \ac{SDMA} operational region \ac{SE} may decrease slightly. Numerical results confirm these findings, with \ac{SE} gains reaching up to 13\% in the \ac{NOMA} regime and reductions of no more than 2\% in the \ac{SDMA} regime. Our findings also revealed that the \ac{SCA}-based iterative algorithm is unable to find feasible \ac{RSMA} precoders in scenarios with high secrecy requirements, even though such solutions exist. Therefore, future research should explore alternative solution methods for the \ac{SE}-maximization problem to evaluate secure \ac{RSMA} systems more reliably under challenging conditions, such as higher numbers of users and antennas.\looseness=-1

%% file: chapters/08_Acknowledgement.tex
\section*{Acknowledgment} \small
\footnotesize{This work was supported by the Federal Ministry of Research, Technology and Space (BMFTR) as part of the 6G-ANNA Project under Grant 16KISK103 and the 6G Platform Project under Grant 16KISK051.}